# Identifying Treatment and Spillover Effects with Control-Based and Forecast-Based Counterfactuals

Viviana Celli*, Augusto Cerqua‡, Guido Pellegrini§

This version: 22/07/2026

**Abstract:** *Spillovers and interference pose fundamental challenges for causal inference, as treatment assigned to one unit may affect the outcome of others, violating the no-interference assumption underlying most empirical strategies. Existing approaches, based on partial interference, exposure mapping, spatial, network, or structural frameworks, typically rely on strong assumptions about interaction structures or require the existence of uncontaminated control units to estimate relevant causal parameters. We revisit this identification challenge within the potential outcomes framework and compare the conditions under which causal effects can be identified using two broad classes of counterfactual methods: control-based counterfactual methods (CBCMs), such as matching and difference-in-differences designs, and forecast-based counterfactual methods (FBCMs), including interrupted time-series and machine learning control methods. We show under which circumstances CBCMs and FBCMs identify average direct and spillover effects. Through simulations and an empirical application, we illustrate the main advantages and limitations of each approach. We show that, in the presence of pervasive or ill-defined spillover effects, CBCMs either cannot be used or entail severe identification concerns, whereas FBCMs can more credibly identify some of the causal parameters of interest, at least in the short term.*

Acknowledgments: We are grateful to Manuela Angelucci, Martin Huber, and Pedro Sant'Anna for their valuable feedback. We also thank participants at the COMPIE Conference, the AISRe Conference, and the International Conference on Economic Statistics for helpful comments and suggestions.

* Department of Statistical Sciences, Sapienza University of Rome, Italy. Email: viviana.celli@uniroma1.it
‡ Department of Social Sciences and Economics, Sapienza University of Rome, Italy. Email: augusto.cerqua@uniroma1.it
§ Department of Social Sciences and Economics, Sapienza University of Rome, Italy. Email: guido.pellegrini@uniroma1.it

## 1. Introduction

Spillovers are one of the main unresolved challenges in modern policy evaluation. When the treatment assigned to one unit affects the outcomes of other units, the counterfactual outcomes required for causal inference become harder to define, because untreated observations may themselves be affected by the intervention. As a result, the standard comparison between treated and untreated units no longer identifies the causal effects of interest. Incorporating spillovers into a policy evaluation framework is inherently complex. For tractability, most empirical applications therefore rely on the no-interference assumption, first formalized by Cox (1958) and later incorporated into Rubin's Stable Unit Treatment Value Assumption (SUTVA) (Rubin, 1974, 1980).[1] In this study, we focus exclusively on the no-interference component of SUTVA, which rules out spillovers, externalities, general equilibrium effects, or any form of treatment diffusion across units. While this assumption provides a tractable basis for defining causal effects, it is often unrealistic in settings characterized by social, spatial, or market interactions (Manski, 2013). As a result, reliance on the no-interference assumption may lead to two severe consequences: (i) failure to identify the magnitude and scope of spillover effects, yielding only a partial assessment of the policy under analysis; and (ii) biased estimates of the average treatment effects of interest. As many policies, particularly in public health, education, innovation, and place-based interventions, are explicitly designed to generate spillovers, failing to account for interference may result in misleading inferences. For instance, Sobel (2006) shows that ignoring relevant spillover effects can lead researchers to conclude that an intervention is beneficial, even when it generates adverse effects for both treated and indirectly exposed units.

The existing literature has mainly addressed this problem by explicitly modeling the interference structure through exposure mappings, social networks, spatial interaction matrices or partial interference assumptions. While these approaches differ substantially, they all share a common feature: they require the researcher to specify, either explicitly or implicitly, how spillovers propagate across units before causal effects can be identified. Unfortunately, the interaction structure is rarely observed and is often only imperfectly understood. Geographic proximity, administrative borders or social networks are only proxies for the true transmission mechanism, which is generally unknown and may itself evolve after the intervention.

A seminal contribution highlighting the empirical consequences of interference is Sobel (2006), who shows that, when interference is present, standard difference-in-means estimators no longer identify interpretable causal parameters, even under random treatment assignment, because they confound direct and spillover effects. To address this issue, Sobel introduces the concept of partial interference, whereby interference is allowed within predefined clusters but ruled out across them. Building on this framework, Hudgens and Halloran (2008) develop formal definitions of causal estimands and estimators for direct, indirect, total, and overall effects under partial interference. They propose a two-stage randomization design that allows both group-level and individual-level treatment assignment to vary. Relatedly, Huber and Steinmayr (2021) extend the analysis of interference to observational settings by developing a framework for separating individual-level treatment effects from spillover effects under selection on observables and difference-in-differences (DiD) designs.

[1] SUTVA consists of two components: (i) the absence of interference, meaning that a unit's potential outcome depends solely on its own treatment assignment; and (ii) the absence of hidden variations in the treatment itself.

A growing literature seeks to operationalize interference by mapping neighbors' treatments into exposure measures that define causal estimands under structured interdependence (Halloran and Struchiner, 1995). A fundamental contribution is provided by Aronow and Samii (2017), who show that causal effects remain well defined under interference by indexing potential outcomes not only to a unit's own treatment status but also to an exposure mapping, namely a function that summarizes how treatment assignments received by other units determine the unit's degree of exposure. This framework generalizes earlier work on partial interference and provides a unified representation of direct, indirect, total and overall causal effects. However, existing approaches typically rely on strong and largely untestable assumptions about the structure of interactions and the correct specification of exposure mappings,[2] which are often based on group membership, spatial proximity, social networks, or saturation-based measures defined by the proportion of treated units within a reference group (Tchetgen Tchetgen and VanderWeele, 2012; Angelucci and Di Maro, 2016; Forastiere et al., 2021; Sävje et al., 2021; Leung, 2022; Di Stefano and Mellace, 2024; Grossi et al., 2025; Debarsy and Le Gallo, 2025).

Beyond the partial interference framework, alternative strategies have emerged to operationalize and model interdependence across units. Spatial econometrics provides an alternative framework by explicitly modeling cross-sectional dependence through spatial lag or spatial Durbin specifications (Anselin, 1988; LeSage and Pace, 2009). In these models, spillovers are treated as structural components of the data-generating process, allowing researchers to estimate interaction effects directly through spatial connectivity matrices. However, these models are primarily designed to fit a spatial structure rather than to identify causal estimands, and causal interpretation is often limited (Gibbons and Overman, 2012; Debarsy and Le Gallo, 2025). Their reliance on exogenously specified interaction matrices, typically based on geographic proximity, may fail to capture true transmission channels, and they remain vulnerable to spatial confounding, endogenous sorting, and common shocks (Bai and Li, 2021; Shi and Lee, 2018). Within this same spatial econometric framework, instrumental variables have been proposed to address endogeneity concerns; however, finding credible instruments for endogenous spatial lags remains generally difficult (Kelejian and Prucha, 2010; Agrawal et al., 2022).

Closely related, network-based frameworks conceptualize interference as arising from explicitly defined relational structures among units. By allowing treatment effects to propagate along network edges, these approaches accommodate heterogeneous and localized spillover effects and offer a richer representation of interaction mechanisms compared to purely spatial approaches (Bramoullé et al., 2009; Jackson, 2010). At the same time, causal interpretation often depends on assumptions about the observability, stability, and exogeneity of the network structure. These assumptions may be demanding in settings where networks are shaped by social or economic selection processes, although a growing econometric literature explicitly models endogenous network formation (Goldsmith-Pinkham and Imbens, 2013; de Paula, 2017). In such cases, endogenous network formation, measurement error in relational data, and network evolution over time can complicate identification. Moreover, separating peer influence from correlated and contextual effects generally requires either strong structural assumptions, prior information on reference groups, or exogenous variation in network exposure (Manski, 1993; Ogburn and VanderWeele, 2017).

[2] Sävje (2024) is an important exception. In randomized experiments, he shows that misspecified exposure mappings can still define estimable expected exposure effects, provided the induced specification errors are weakly dependent and standard design conditions hold. These effects, however, are defined relative to the researcher's exposure labels and the implemented assignment design, not necessarily to the true interference structure. Leung (2024) and Auerbach et al. (2024) raise important concerns about their causal interpretation and policy relevance under misspecification, highlighting an interpretability–estimability trade-off in the choice of estimand: effects that are easier to estimate under weak assumptions may be harder to interpret causally, while more clearly interpretable causal effects typically require stronger assumptions about the interference structure. Similarly, Mealli and Viviens (2025) show that, under unknown interference, the conventional DiD estimator identifies a contrast between the total effect on treated units and the spillover effect on control units, rather than either effect separately.

Finally, structural modeling approaches embed interference mechanisms within fully specified behavioral or equilibrium models (Heckman and Vytlacil, 2007; Aguirregabiria and Mira, 2010). By explicitly modeling decision-making processes, strategic interactions, or market equilibria, these models provide theoretically coherent interpretations of spillover effects and allow counterfactual simulations under alternative policy scenarios (Blundell and Costa Dias, 2009). Meanwhile, their causal interpretation depends on the credibility of the underlying behavioral assumptions, functional-form choices, and restrictions used to identify the relevant structural parameters (Lewbel, 2019; Debarsy and Le Gallo, 2025). When these assumptions are difficult to validate empirically, inference on spillover effects may become highly sensitive to model specification.

Overall, existing approaches typically require an explicit specification of the interaction structure or rely on strong structural assumptions. Point identification of policy-relevant causal effects remains difficult under pervasive or poorly understood interference without restrictive design or modeling conditions. Moreover, much of this literature constructs counterfactual outcomes through comparison units that are assumed to be unaffected by the intervention. An alternative approach uses time-series or panel-data dynamics to construct counterfactuals through forecasts (Box and Tiao, 1975; Menchetti et al., 2023; Cerqua et al., 2024; Botosaru et al., 2026). To our knowledge, however, the role of forecast-based counterfactuals for identifying treatment and spillover effects has received no explicit attention. The objective of this paper is not to propose another model of interference, but to show that, for total policy effects, explicit modeling of the interference mechanism is not even required. By replacing cross-sectional comparisons with forecast-based counterfactuals, causal effects can be identified under weaker assumptions about the propagation of spillovers.

Our methodological approach presents treatment and spillover effects in the potential outcomes framework, as it strikes a balance between model complexity, credibility of the identifying assumptions, and the ability to recover the causal estimands of interest. We characterize identification under different interference scenarios for two broad classes of counterfactual methods: control-based counterfactual methods (CBCMs), such as difference-in-differences (DiD) and matching estimators, and forecast-based counterfactual methods (FBCMs), including interrupted time-series (ITS) designs and the machine learning control method (MLCM). By relying on model-based forecasts of Y, rather than on outcomes observed for untreated units, FBCMs can estimate counterfactual outcomes even when untreated units are themselves exposed to spillovers, thereby relaxing the need for policy-unaffected units. This flexibility, however, comes at the cost of assumptions about the temporal stability and predictability of untreated outcomes, which must be assessed case by case and may be difficult to sustain for long-term effects.

In summary, this paper makes three main contributions to the causal inference literature. First, we show that aggregate effects defined relative to the no-treatment/no-exposure counterfactual can be identified without specifying an exposure mapping, whereas exposure-specific effects continue to require information on the exposure structure. Second, we develop a unified identification framework that accommodates both control-based and forecast-based counterfactual approaches within the potential outcomes setting. Third, we formally demonstrates that FBCMs can identify relevant causal estimands even when interference is pervasive and all units are affected directly or indirectly by the treatment (no clean control), a setting in which CBCMs are not suitable without additional strong assumptions.

To examine how CBCMs and FBCMs perform when the exposure mapping is misspecified, we conduct a simulation study based on the actual geography of U.S. counties. In each Monte Carlo replication, 200 counties are randomly assigned to treatment, while spillovers propagate according to queen contiguity and increase with the number of treated neighboring counties. We compare a benchmark in which the researcher observes the true exposure mapping with two misspecified mappings that classify counties according to whether their centroids lie within 30 or 60 km of a treated county. The balanced panel contains five pre-treatment periods and one post-treatment period, and outcomes exhibit systematic level differences, persistent covariates, nonlinearities, and autoregressive adjustment while preserving parallel untreated-outcome trends in expectation. The simulations assess the performance of a doubly robust control-based estimator and two forecast-based approaches. They show how mapping errors contaminate apparently clean controls and alter the composition of exposure-specific target groups. Forecast-based unit-level counterfactuals do not depend on the assumed mapping, although estimates of exposure-specific effects remain sensitive to how those effects are aggregated, whereas the full-population policy effect does not.

In the empirical application, we revisit the natural experiment studied by Di Tella and Schargrodsky (2004) on police protection and car theft in Buenos Aires after the 1994 terrorist attack against the A.M.I.A. Jewish center. This setting provides a natural testing ground for our framework: police protection may reduce theft in directly protected blocks, but it may also displace crime or generate deterrence spillovers in surrounding areas, making uncontaminated controls difficult to define. We compare control-based DiD estimates based on distance-defined exposure groups with forecast-based estimates from the MLCM. The results highlight the main empirical trade-off analyzed in the paper: CBCMs are informative when clean controls can be credibly identified, whereas FBCMs remain useful when the spatial structure of spillovers is uncertain.

The rest of the paper proceeds as follows: Section 2 introduces the potential outcomes framework in the presence of spillovers, Section 3 presents the proposed control-based and forecast-based counterfactual methodologies, Section 4 reports the simulation design and results, Section 5 discusses the empirical application, and Section 6 concludes.

## 2. The potential outcomes framework with spillovers

Consider a balanced panel of units $i = 1, \dots, N$ observed over periods $t = 1, \dots, T_0, \dots, T$. Let $T_0$ denote the common treatment adoption date.[3] All units are untreated for $t < T_0$, while a subset of units becomes treated from $T_0$ onward. Let $D_i \in \{0,1\}$ be a time-invariant treatment assignment indicator, with realized treatment status

$$D_{it} = D_i \mathbf{1}\{t \geq T_0\}.$$

We explicitly allow for non-random treatment assignment. In particular, treated and untreated units may differ in outcome levels prior to treatment, while sharing a common pre-treatment trend. This would be an ideal setting for a difference-in-differences design, except for the potential presence of spillover effects.

[3] For simplicity, we use notation based on a common treatment date. The same framework can be readily extended to cases of staggered treatment adoption.

We also allow for interference among units. Let $W = (w_{ij})$ be a time-invariant spatial or network weight matrix, with $w_{ij}$ measuring the strength of interaction between units $i$ and $j$, and $w_{ii} = 0$ ruling out self-interaction. The matrix allows interactions between both treated and untreated units.

Define post-treatment exposure to treated neighbors as

$$S_i = \sum_{j \neq i} w_{ij} D_j = (W\mathbf{D})_i,$$

where $\mathbf{D}$ is the vector of time-invariant treatment assignments. Let $\mathcal{S}$ denote the support of $S_i$ in the population, and let $\mathcal{S}_1$ and $\mathcal{S}_0$ denote the supports of exposure among treated and untreated units, respectively. Exposure summarizes the intensity of unit $i$'s exposure to treatment received by other units and provides an exposure mapping in the sense of Aronow and Samii (2017). Exposure $S_i$ is time invariant because it is determined by the fixed cross-sectional configuration of treated units and by the spatial or network structure encoded in $W$.[4] However, its effect on outcomes is realized only in periods $t \geq T_0$, when the treatment becomes active. We therefore define the effective exposure entering the observed outcome as

$$S_{it} = S_i \mathbf{1}\{t \geq T_0\}.$$

In the presence of interference, the potential outcome of unit $i$ at time $t$ is written as

$$Y_{it}(d, s), \qquad d \in \{0,1\},\ s \in \mathcal{S},$$

where $d$ denotes own treatment status and $s$ denotes exposure to treated neighbors. The observed outcome is

$$Y_{it} = Y_{it}(D_{it}, S_{it}).$$

This formulation allows spillovers to affect both untreated and treated units and nests the standard no-interference framework as a special case. Interference may arise through two distinct, potentially coexisting channels. The first is treated-to-untreated interference, whereby treatment assigned to treated units affects the outcomes of untreated units, making some or all untreated units indirectly exposed to the treatment. The second is within-treated interference, whereby treatment assigned to one treated unit affects the outcomes of other treated units, so that treated units may be affected not only by their own treatment status but also by the treatment status of other treated units. Importantly, this second form of interference complicates both the interpretation and identification of average treatment effects, even when attention is restricted to treated units.[5]

The general framework above encompasses four empirically relevant scenarios.

---

[4] We focus on the realized treatment assignment vector observed in the data. In the presence of interference, exposure status and exposure intensity are functions of this realized assignment configuration. Hence, different realizations of $\mathbf{D}$ could imply different exposure mappings, different shares of treated and exposed units, and different aggregate estimands. The effects studied here are therefore conditional on the observed realization of the policy.

[5] Even in the limiting case where all units are treated, spillover effects may still arise through interactions among treated units. For example, in studies of excess mortality during the COVID-19 pandemic, areas with limited intensive care capacity were compelled to transfer patients to neighboring hospitals, potentially inducing indirect increases in mortality in the receiving facilities as well (see Cerqua et al., 2021).

Case (i): No spillovers.

If outcomes do not depend on exposure,

$$Y_{it}(d,s) = Y_{it}(d,0) \quad \text{for all } s \in \mathcal{S},$$

the model collapses to the standard no-interference framework. The primary causal estimand of interest is the Average Treatment Effect on the Treated (ATT):[6]

$$ATT_t \equiv E[Y_{it}(1,0) - Y_{it}(0,0) \mid D_i = 1], \qquad t \geq T_0.$$

In such a scenario, there is no need to estimate average spillover effects, as they are absent by assumption.

In Section 3, we show that both CBCMs and FBCMs can point identify $ATT_t$ under their respective identifying assumptions.

Case (ii.a): Spillovers affecting a subset of units (selective interference).

In many applications, spillovers affect only a subset of units. To capture selective interference, we distinguish between exposed units $(S_i > 0)$ and unexposed units $(S_i = 0)$. By definition, spillover effects on outcomes are realized only for $t \geq T_0$, when the treatment becomes active. This setting combines selective exposure with partial interference in the sense of Sobel (2006), provided that the interaction matrix $W$ is block diagonal with respect to the predefined clusters, so that spillovers may occur within, but not across, clusters. Information is assumed to be available on which untreated units are exposed, on their exposure intensity $s$, and on which units remain unexposed. Importantly, spillovers may affect both treated and untreated units.

Within this framework, several causal estimands can be defined at a given exposure level $s$. Let us start by considering the subgroup of treated units. The direct effect at exposure level $s$ for these units, also known as the ATT conditional on exposure $s$, is defined as

$$DE_t(s) \equiv ATT_t(s) \equiv E[Y_{it}(1,s) - Y_{it}(0,s) \mid D_i = 1, S_i = s], \qquad s \in \mathcal{S}_1,$$

where $\mathcal{S}_1$ denotes the support of observed exposure levels among treated units. $DE_t(s)$ captures the effect of own treatment while holding exposure fixed at level $s$.

Treated units may also be subject to indirect spillover effects from other treated units. In this case, we can define the indirect effect on the treated as

$$IE_t(s) \equiv E[Y_{it}(1,s) - Y_{it}(1,0) \mid D_i = 1, S_i = s], \qquad s \in \mathcal{S}_1,$$

measuring the effect of exposure among treated units while holding own treatment status fixed at $d = 1$.

The total effect at exposure level $s$ for the treated is then given by

$$TE_t(s) \equiv E[Y_{it}(1,s) - Y_{it}(0,0) \mid D_i = 1, S_i = s], \qquad s \in \mathcal{S}_1.$$

[6] For simplicity, we abstract from treatment-effect heterogeneity driven by interactions between treatment and unit-specific characteristics (e.g., age or gender), typically captured by conditional average treatment effects (CATEs). Such heterogeneity can be incorporated straightforwardly within our framework.

To decompose this total effect, define the direct effect in the absence of exposure for the same subgroup of treated units as

$$DE_t^0(s) \equiv E[Y_{it}(1,0) - Y_{it}(0,0) \mid D_i = 1, S_i = s], \qquad s \in \mathcal{S}_1.$$

Then the total effect can be written as the sum of the direct effect in the absence of exposure and the indirect effect among treated units:

$$TE_t(s) = DE_t^0(s) + IE_t(s).$$

Hence, the total effect combines the effect of own treatment when exposure is set to zero and the additional effect of exposure among treated units.[7]

**Proposition 1.** Fix a period $t \geq T_0$ and an exposure level $s \in \mathcal{S}_1$. Recall the definitions

$$DE_t^0(s) \equiv E[Y_{it}(1,0) - Y_{it}(0,0) \mid D_i = 1, S_i = s],$$
$$IE_t(s) \equiv E[Y_{it}(1,s) - Y_{it}(1,0) \mid D_i = 1, S_i = s],$$
$$TE_t(s) \equiv E[Y_{it}(1,s) - Y_{it}(0,0) \mid D_i = 1, S_i = s].$$

Then

$$TE_t(s) = DE_t^0(s) + IE_t(s).$$

The proof is reported in Appendix A.

Averaging the total effect over the realized exposure distribution among treated units yields the Average Total Effect on the Treated (ATOT),

$$ATOT_t \equiv E[Y_{it}(1, S_i) - Y_{it}(0,0) \mid D_i = 1],$$

that is, the expectation averages $TE_t(s)$ over the empirical distribution of exposure on $\mathcal{S}_1$. $ATOT_t$ captures the overall impact of the treatment on treated units, inclusive of the indirect effects they experience.[8]

The indirect spillover effect on untreated exposed units is defined as

$$SE_t(s) \equiv E[Y_{it}(0,s) - Y_{it}(0,0) \mid D_i = 0, S_i = s], \qquad s \in \mathcal{S}_0,\ s > 0,$$

measuring the impact of exposure in the absence of direct treatment and where $\mathcal{S}_0$ denotes the support of observed exposure levels among untreated units.[9]

---

[7] When $s = 0$, treated units are not exposed to spillovers, and the total effect coincides with the ATT conditional on treated units having zero exposure. It coincides with the unconditional $ATT_t$ defined in Case (i) only if all treated units are unexposed or treatment effects are homogeneous across exposure groups.

[8] One could also define average direct effects and average indirect effects by averaging ($DE_t(s)$ and $IE_t(s)$), respectively, over the relevant exposure distribution. We do not report these additional aggregate estimands here to avoid further burdening the notation.

[9] We use different notation for indirect spillover effects on treated and untreated exposed units. $IE_t(s)$ refers to the effect of exposure among treated units, holding own treatment status fixed at $D_i = 1$, while $SE_t(s)$ refers to the spillover effect on untreated exposed units, holding own treatment status fixed at $D_i = 0$. This distinction avoids conflating within-treated interference with spillovers affecting untreated units, as the two objects require different identifying assumptions. It also reflects the fact that we do not restrict the effect of exposure to be the same when it is combined with direct treatment and when it occurs in the absence of direct treatment.

These exposure-specific estimands can be aggregated to obtain policy-relevant average effects. Averaging the spillover effect over the realized distribution of exposure among affected untreated units yields the Average Spillover Effect on the Exposed Untreated (ASEU),

$$ASEU_t \equiv E[Y_{it}(0, S_i) - Y_{it}(0,0) \mid D_i = 0, S_i > 0],$$

which integrates $SE_t(s)$ over the distribution of $S_i$ on $\{s \in \mathcal{S}_0\colon\ s > 0\}$.

Finally, one may also be interested in the overall average impact of the treatment on the entire population, not only on treated units or on exposed untreated units. Following the logic of the overall causal effect proposed by Hudgens and Halloran (2008), we define the Population Average Policy Effect (PAPE) as

$$PAPE_t \equiv E[Y_{it}(D_i, S_i) - Y_{it}(0,0)].$$

This estimand compares the realized treatment and exposure configuration with a counterfactual scenario in which no unit is treated and no unit is exposed to spillovers. From a policy perspective, this is arguably the most relevant causal estimand, as it encompasses the average impact of the treatment on the whole population.[10] This population-level estimand is related to the welfare-based perspective of Viviano (2025), who studies policy targeting under network interference. In our setting, $PAPE_t$ is the natural evaluation counterpart of that perspective, as it measures the average effect of the realized policy configuration on the whole population.

Under the selective-interference structure considered here, the population can be partitioned into three mutually exclusive groups: directly treated units, untreated exposed units, and untreated unexposed units. Directly treated units are kept as a single group, despite possible heterogeneity in their exposure levels, because $ATOT_t$ averages total effects over the realized exposure distribution among them. Therefore, the overall effect can be written as the weighted average of the average total effect on treated units, the average spillover effect on exposed untreated units, and the zero effect on untreated unexposed units:$PAPE_t = \Pr(D_i = 1)\, ATOT_t + \Pr(D_i = 0, S_i > 0)\, ASEU_t + \Pr(D_i = 0, S_i = 0) \cdot 0.$

Although the last term is equal to zero in light of the zero impact of the treatment on untreated unexposed units, it should not be ignored conceptually. Untreated and unexposed units are part of the population over which the average is computed, and therefore their presence dilutes the population-level treatment effect toward zero.

Equivalently, if we define the average policy effect among directly or indirectly exposed units (ATTE) as

$$ATTE_t \equiv E[Y_{it}(D_i, S_i) - Y_{it}(0,0) \mid D_i = 1 \cup S_i > 0],$$

then the PAPE can be written as

$$PAPE_t = \Pr(D_i = 1 \cup S_i > 0)\, ATTE_t.$$

These estimands are well defined once potential outcomes are indexed by both treatment and exposure.

As discussed in Section 3, under selective interference and perfect knowledge of the exposure mapping, the identification of $TE_t(s)$, $ATOT_t$, $SE_t(s)$, $ASEU_t$, $ATTE_t$, and $PAPE_t$ can be achieved under a reasonable set of assumptions by both CBCMs and FBCMs. At the same time, identifying $DE_t(s)$ and $IE_t(s)$ in practice may be more challenging, as it requires sufficient overlap between exposed and unexposed treated units, as well as between treated and untreated units at comparable exposure levels.

[10] We use the label PAPE to emphasize that the estimand refers to the average effect of the realized policy configuration, including both direct treatment and spillover exposure, rather than to the standard no-interference population average treatment effect.

Case (ii.b): Latent selective interference.

A conceptually distinct situation arises when spillovers affect only a subset of units, but the researcher cannot observe which units are exposed. In this setting, untreated and unexposed units still exist in the population, and exposure remains heterogeneous across units, but exposure status is unobserved or measured with substantial error. Formally, there exist units satisfying $S_i = 0$ and units satisfying $S_i > 0$, but the researcher cannot reliably distinguish between them. This situation may arise when spillovers diffuse through unobserved social, economic, or geographic networks, or when the available exposure measure is affected by substantial measurement error.

We are interested in the same causal estimands as in Case (ii.a). However, as shown in Section 3, these estimands cannot generally be point identified using CBCMs. Consequently, the existence of clean controls is not sufficient for point identification: researchers must also be able to correctly identify which units are uncontaminated.[11]

Conversely, under latent selective interference, FBCMs can still point identify aggregate total effects, such as $ATOT_t$ and $PAPE_t$, while it is no longer possible to point identify exposure-specific or exposure-status-dependent objects such as $TE_t(s)$, $SE_t(s)$, $ASEU_t$, or $ATTE_t$.

Case (iii): Spillovers affecting all units.

In some settings, such as those characterized by general equilibrium effects or highly interconnected systems, spillovers may affect all units, both treated and untreated, so that $S_i > 0$ for all $i$. Suppose a large infrastructure program improves public transportation in one part of a metropolitan area. Even if the policy is geographically targeted, it may increase overall citywide accessibility, affect commuting patterns, and change housing demand across all neighborhoods. As a result, rents and property values may adjust not only in treated areas but also in untreated ones, because households reallocate in response to new relative prices and commuting costs. In this case, all units are indirectly exposed to the treatment through market-clearing mechanisms in the housing and labor markets. Hence, $S_i > 0$ for all $i$ and $PAPE_t = ATTE_t$.[12]

As shown in Section 3, reduced-form approaches based on cross-unit comparisons including CBCMs break down in this setting, since all untreated units are themselves affected by spillovers and no control group remains unaffected by the treatment, provided that the available data cover only the metropolitan area under study.[13]

[11] Auerbach et al. (2024) note that, even when the exposure mapping is misspecified, an exposure-based comparison may still be useful for the narrower purpose of testing the null hypothesis that all spillover effects are zero. In this case, a statistically significant nonzero misspecified exposure effect provides evidence against the absence of spillovers, although it cannot generally be interpreted as a policy-relevant measure of the magnitude or spatial structure of spillover effects.

[12] Cerqua et al. (2024) study an extreme version of Case (iii), in which all units are simultaneously exposed to treatment. In such circumstances, CBCMs cannot be used to identify the causal estimands of interest because they require untreated observations to identify the counterfactual path.

If the exposure mapping is known and there is sufficient common support in exposure between treated and untreated units, CBCMs may still identify a more limited causal object: the direct effect on the treated at a given exposure level, $DE_t(s)$, under a parallel-trends assumption conditional on $s$; see Xu (2025) for an example. This estimand compares treated and untreated units facing the same exposure level and therefore does not recover total effects relative to a no-treatment/no-exposure counterfactual, nor spillover effects on untreated exposed units. Under the same known-exposure setting, FBCMs can identify causal estimands defined relative to $Y_{it}(0,0)$, such as $TE_t(s)$, $ATOT_t$, $SE_t(s)$, $ASEU_t$, and $PAPE_t$, but they do not generally identify $DE_t(s)$ or $IE_t(s)$ without additional structure. Since all units are exposed in this case, the population of directly or indirectly exposed units coincides with the entire population. Therefore, $ATTE_t$ coincides with $PAPE_t$ and can be identified whenever $PAPE_t$ is identified.

If the exposure mapping is instead unknown, in Section 3 we show that CBCMs cannot identify any of the causal estimands defined here, while FBCMs can still point identify aggregate effects such as $ATOT_t$, $ASEU_t$, and $PAPE_t$.

## 3. The control-based and forecast-based counterfactual approaches

This section describes two broad classes of approaches for constructing counterfactual outcomes in panel data settings: control-based counterfactual methods (CBCMs) and forecast-based counterfactual methods (FBCMs). The key distinction between the two lies in how the post-treatment counterfactual potential outcomes are recovered. Control-based approaches rely on cross-unit comparisons, whereas forecast-based approaches recover counterfactual outcomes from the pre-treatment dynamics of the units for which counterfactual outcomes are needed, using information on their past outcomes and covariates. As we show below, this distinction is crucial for identification, especially in the presence of interference.

### 3.1. Control-based counterfactual approaches

CBCMs recover the counterfactual outcome $Y_{it}(0,0)$ by leveraging outcomes of untreated units that are assumed to be unaffected by the policy. Difference-in-differences (DiD) is the canonical example. Identification relies on assumptions linking observed untreated outcomes to the untreated and unexposed potential outcome. We distinguish between four interference regimes corresponding to the interference structures described in Section 2.

#### 3.1.1. CBCMs assumptions

Assumption CBCM1 (No anticipation). For all $t < T_0$,

$$Y_{it}(d,s) = Y_{it}(0,0) \qquad \forall d \in \{0,1\}, \;\; s \in \mathcal{S}$$

Pre-treatment outcomes are therefore realizations of the untreated and unexposed potential outcome. This assumption ensures that pre-treatment outcomes provide valid information about the untreated outcome process, as it guarantees that pre-treatment data are not contaminated by behavioral or equilibrium responses to the anticipated policy.

[13] If comparable untreated and unaffected units are observed outside the metropolitan area under study, for example in other cities not affected by the policy, the setting becomes analogous to Case (ii.a).

Assumption CBCM2 (No interference with control units). For all untreated units,

$$Y_{it}(0,s) = Y_{it}(0,0) \qquad \forall i\colon D_i = 0, \forall s \in S_0,$$

so that observed outcomes for untreated units coincide with $Y_{it}(0,0)$, i.e., with the untreated and unexposed potential outcome.

When interference affects some untreated units, Assumption CBCM2 is violated: spillovers contaminate untreated outcomes, breaking the link between observed untreated outcomes and $Y_{it}(0,0)$ when $t \geq T_0$, and thereby invalidating the cross-unit comparisons that underpin any control-based approach. However, some relevant causal estimands may still be identified by relying on the following weaker version of Assumption CBCM2.

Assumption CBCM2_v2 (Existence of clean controls). There exists a non-empty subset of untreated units such that,

$$Y_{it} = Y_{it}(0,0) \qquad \forall i\colon D_i = 0, S_i = 0.$$

Moreover, the researcher must observe which units satisfy $S_i = 0$.

Then, assumption CBCM2_v2 has two components:

(i) existence of untreated and unexposed units;

(ii) observability of such units by the researcher.

When condition (i) holds but condition (ii) fails, we obtain the latent selective interference setting introduced in Section 2 (Case ii.b).

Because different estimands use different comparison groups, we distinguish three parallel-trends conditions.

Assumption CBCM3(a) (Parallel trends: treated units versus the relevant valid-control group). In the absence of treatment and spillovers, the untreated potential outcomes of treated units and those of the relevant valid-control units would have evolved in parallel. The relevant control group depends on the interference regime. Formally:

Case (i): No spillovers. For $t \geq T_0, t' < T_0$,

$$E[Y_{it}(0,0) - Y_{it'}(0,0) \mid D_i = 1] = E[Y_{it}(0,0) - Y_{it'}(0,0) \mid D_i = 0].$$

Case (ii.a): Selective spillovers. For $t \geq T_0, t' < T_0$, and for any exposure level $s \in S_1$,

$$E[Y_{it}(0,0) - Y_{it'}(0,0) \mid D_i = 1, S_i = s] = E[Y_{it}(0,0) - Y_{it'}(0,0) \mid D_i = 0, S_i = 0].$$

In Case (i), all untreated units constitute valid controls because their outcomes are unaffected by exposure. In Case (ii.a), the valid-control group must instead be restricted to untreated and unexposed units. Condition (ii.a) imposes parallel untreated-outcome trends within each exposure stratum among treated units. Under the remaining identifying assumptions, it supports identification of $TE_t(s)$; averaging over the realized exposure distribution among treated units yields identification of $ATOT_t$.

Assumption CBCM3(b) (Parallel trends: exposed untreated units vs clean controls). For any positive exposure level $s \in S_0$ and for $t \geq T_0, t' < T_0$,

$$E[Y_{it}(0,0) - Y_{it'}(0,0) \mid D_i = 0, S_i = s] = E[Y_{it}(0,0) - Y_{it'}(0,0) \mid D_i = 0, S_i = 0].$$

This condition ensures that clean controls provide a valid counterfactual trend for exposed untreated units.

Assumption CBCM3(c) (Exposure-specific parallel trends). For any exposure level $s \in \mathcal{S}_1 \cap \mathcal{S}_0$ with common support among treated and untreated units, and for $t \geq T_0, t' < T_0$,

$$E[Y_{it}(0,s) - Y_{it'}(0,0) \mid D_i = 1, S_i = s] = E[Y_{it}(0,s) - Y_{it'}(0,0) \mid D_i = 0, S_i = s].$$

This condition requires that, conditional on the same exposure level $s$, treated and untreated units would have followed parallel trends in the potential outcome corresponding to no own treatment and exposure level $s$. It is the additional parallel-trends condition needed when the comparison is made between treated and untreated units facing the same exposure level.

The no-anticipation and clean-control conditions are common to CBCMs, including designs relying on random treatment assignment, whereas CBCM3(a), CBCM3(b) and CBCM3(c) are specific to DiD-type estimators.

These parallel-trends assumptions are primarily used to identify estimands defined relative to $Y_{it}(0,0)$, such as $TE_t(s)$, $ATOT_t$, $SE_t(s)$, $ASEU_t$, and $PAPE_t$. By contrast, $DE_t(s)$ and $IE_t(s)$ require additional overlap and exposure-specific parallel-trends conditions, because they involve counterfactual outcomes at nonzero exposure levels.

**Remark (Latent Selective Interference).** Assumptions CBCM3(a)-CBCM3(c) are formulated in terms of exposure status and may therefore hold at the population level even when exposure is not observed by the researcher. In the latent selective interference setting, however, exposure status cannot be used to partition observations into clean controls and exposed units. As a result, even if these assumptions remain valid features of the data-generating process, they cannot be implemented empirically for identification. The failure of CBCMs in Case (ii.b) therefore stems not from a violation of parallel trends, but from the inability to identify the exposure regimes required by these assumptions.

The credibility of these assumptions depends on the institutional context and the structure of interactions across units. Assumption CBCM1 may fail when agents adjust their behavior in anticipation of the policy. In such cases, the researcher should treat as the last pre-treatment period the final period in which it is reasonable to assume that agents could not yet respond to the forthcoming policy change. Assumptions CBCM2 and CBCM2_v2 require that all or some untreated units remain unaffected by spillovers and that the researcher can correctly identify them, conditions that may be implausible in highly interconnected settings. Finally, the parallel-trends assumptions (CBCM3(a)-CBCM3(c)) are more credible when treated and comparison groups share similar pre-treatment dynamics and are exposed to common shocks, and less credible when treatment assignment or exposure is correlated with differential underlying trends.

### 3.1.2. The difference-in-differences design in the presence of interference

A common empirical strategy for studying spillovers in observational settings is to combine a control-based causal design with an explicit exposure mapping. The researcher first identifies the directly treated units and then defines untreated units that may be indirectly exposed to the treatment, typically based on geographic proximity, adjacency, network links, or other theoretically motivated measures of exposure. Units that are sufficiently far from the treatment source, or otherwise plausibly disconnected from it, are used as clean controls. This leads to a spatial difference-in-differences design, also known as a difference-in-differences design with interference, in which the post-treatment change in outcomes for exposed untreated units is compared with the corresponding change for untreated and unexposed units (see Butts, 2023). This method is described in Appendix B. More flexible versions estimate separate coefficients for different exposure bins, such as distance rings around the treated location, allowing the researcher to assess how effects decay with exposure intensity. The key identifying assumptions are that, absent treatment and spillovers, exposed and unexposed untreated units would have followed parallel trends, and that the clean-control group is not itself affected by the treatment.

Recent applications illustrate how this logic is implemented in practice. Garin and Rothbaum (2025) adopt a control-based design in which treated units are counties where large government-financed plants were built, while clean controls are defined as similarly populated counties located outside major manufacturing centers and not adjacent to treated counties. Since plant construction may generate spillovers across county borders, adjacent counties are excluded from the clean control group and treated as potentially exposed units. Using the clean controls as the relevant counterfactual group, the authors estimate both the average treatment effect on treated counties and the average spillover effect on adjacent counties.

Bradt and Aldy (2025) study the housing-market effects of US Army Corps of Engineers levee construction. Their design distinguishes between homes located inside levee-protected areas, which receive direct protection benefits, and homes outside protected areas but close to waterways, which may be exposed to adverse spillover flood risks. They estimate difference-in-differences models using repeat residential property transactions and compare price changes around levee construction across protected, spillover-exposed, and unaffected parcels. They also vary the distance-to-water bandwidth used to define spillover exposure, thereby assessing the spatial reach of spillover effects.

Diamond and McQuade (2019) provide another prominent example in the context of affordable housing. They study Low-Income Housing Tax Credit developments and estimate their spillover effects on nearby neighborhoods using a nonparametric difference-in-differences style estimator. Their approach generalizes the standard "ring" method, which compares properties close to the treated site with properties located in an outer control ring. Rather than imposing a single treatment radius, they recover how house-price effects vary flexibly with distance from the LIHTC site and time since development, thereby estimating the spatial reach of the spillover effect.

Together, these examples show that CBCMs can be adapted to spillover settings when the researcher can credibly define exposure regimes. Identification depends not only on standard parallel-trends-type assumptions, but also on the credibility of the exposure mapping used to separate indirectly exposed units from clean controls. Clean controls must be comparable to treated and exposed units in the absence of the intervention while also being plausibly unaffected by the treatment, so that they provide valid counterfactuals without being contaminated by spillovers. This requirement involves a clear trade-off: defining exposure too broadly may unnecessarily exclude unaffected units from the control group, reducing statistical power and the pool of comparable controls, whereas defining exposure too narrowly may leave contaminated units among the clean controls, leading to biased estimates.

### 3.1.3. CBCMs identification results

We now establish identification under the four interference regimes introduced in Section 2.

Case (i): No spillovers.

When potential outcomes do not depend on exposure,

$$Y_{it}(d,s) = Y_{it}(d,0) \qquad \text{for all } s \in \mathcal{S},$$

**Proposition CBCM1.** Under Assumptions CBCM1, CBCM2, and CBCM3(a), the standard DiD estimator identifies $ATT_t$.

Case (ii.a): Spillovers affecting a subset of units (selective interference).

Under this selective-interference structure, untreated and unexposed units, i.e. units with $D_i = 0$ and $S_i = 0$, are the natural candidate control group for recovering the counterfactual outcome $Y_{it}(0,0)$, namely the outcome in the absence of both direct and spillover effects.

When spillovers affect only a subset of units and untreated unexposed units can be observed, CBCMs can still be applied by restricting the comparison group to untreated and unexposed units. In this setting, under no anticipation and the existence and observability of clean controls, DiD-type estimators can identify $TE_t(s)$ and $ATOT_t$ under CBCM3(a), and $SE_t(s)$ and $ASEU_t$ under CBCM3(b). The identification of $ATTE_t$ and $PAPE_t$ then follows from the identification of $ATOT_t$ and $ASEU_t$, together with the relevant population shares.

**Proposition CBCM2.** Fix $t \geq T_0$ and $s \in \mathcal{S}_1$. Under Assumptions CBCM1, CBCM2_v2, and CBCM3(a), a DiD estimator comparing treated units with exposure $s$ to untreated and unexposed units identifies $TE_t(s)$.

**Corollary CBCM1.** A practically relevant special case is the coincidence of the total effect with the direct effect for unexposed treated units, that is, when $s = 0$:

$$TE_t(0) = DE_t(0).$$

**Proposition CBCM3.** Under Assumptions CBCM1, CBCM2_v2, and CBCM3(a), DiD comparing all treated units to untreated unexposed units identifies $ATOT_t$.

**Proposition CBCM4.** Fix $t \geq T_0$ and a positive exposure level $s$ with $s \in \mathcal{S}_0$. Under Assumptions CBCM1, CBCM2_v2, and CBCM3(b), DiD comparing exposed untreated units with exposure $s$ to unexposed untreated units identifies $SE_t(s)$.

**Proposition CBCM5.** If the conditions of Proposition CBCM4 hold for all relevant positive exposure levels, CBCMs identify $ASEU_t$ by averaging $SE_t(s)$ over the realized distribution of exposure among untreated exposed units.

**Corollary CBCM2.** If $ATOT_t$ and $ASEU_t$ are identified and the relevant population shares are observed, then CBCMs also identify $ATTE_t$ and $PAPE_t$.

Finally, $DE_t(s)$ and $IE_t(s)$ require additional conditions. Identifying $DE_t(s)$ requires common support between treated and untreated units at the same exposure level and Assumption CBCM3(c). Identifying $IE_t(s)$ is more demanding, because it requires recovering $Y_{it}(1,0)$, which, in practice, requires overlap between exposed and unexposed treated units and a parallel-trends condition across exposure strata. Absent these conditions, these estimands are not generally point identified by CBCMs.

Case (ii.b): Latent selective interference.

When untreated and unexposed units exist but exposure status is unobserved, CBCMs cannot generally point identify the causal estimands of interest because the researcher cannot isolate the clean-control group required to recover $Y_{it}(0,0)$. The failure is therefore not necessarily due to the absence of clean controls in the population, but to the inability to identify them in the observed data.

**Proposition CBCM6.** Suppose untreated unexposed units exist but exposure status is not observed by the researcher. Then CBCMs cannot generally point identify $TE_t(s)$, $ATOT_t$, $SE_t(s)$, $ASEU_t$, $ATTE_t$, or $PAPE_t$.

Case (iii): Spillovers affecting all units.

When all units are exposed to spillovers, no untreated and unexposed units are available within the observed population. Hence, untreated outcomes no longer provide observations of $Y_{it}(0,0)$, and cross-unit comparisons cannot recover the no-treatment/no-exposure counterfactual.

**Proposition CBCM7.** If all units are exposed to spillovers post-treatment, so that no unit satisfies $D_i = 0$ and $S_i = 0$, CBCMs cannot generally point identify estimands defined relative to $Y_{it}(0,0)$, including $TE_t(s)$, $ATOT_t$, $SE_t(s)$, $ASEU_t$, $ATTE_t$, or $PAPE_t$.

**Proposition CBCM8.** Suppose that the exposure mapping is known and that, for some $s$, there is sufficient common support between treated and untreated units at the same exposure level. Under assumption CBCM3(c), a DiD estimator comparing treated and untreated units with $S_i = s$,

$$\Delta_t^{DE}(s) = E\left[Y_{it} - Y_{i,T_0-1} \mid D_i = 1, S_i = s\right] - E\left[Y_{it} - Y_{i,T_0-1} \mid D_i = 0, S_i = s\right],$$

identifies the direct effect $DE_t(s)$.

The identification proofs presented in Section 3.1.3 are reported in Appendix C.

### 3.2. Forecast-based counterfactual approaches

Forecast-based counterfactual methods (FBCMs) recover the no-treatment, no-exposure potential outcome by extrapolating from pre-treatment outcome and covariate dynamics rather than from cross-unit comparisons. This is a model-based counterfactual approach: counterfactual outcomes are constructed through an explicit statistical model of the baseline data-generating process, and identification relies on the stability and predictability of that process. We use the Machine Learning Control Method (MLCM) (Cerqua et al., 2024) as a leading example of an FBCM. Other approaches, such as interrupted time-series analysis, belong to the same broad family when they recover counterfactual outcomes from pre-treatment dynamics rather than from post-treatment controls.

The core idea of MLCM is to treat counterfactual construction as a forecasting problem. Using only pre-treatment observations, the method estimates a flexible prediction rule for the evolution of outcomes under no treatment and no exposure. Let $\mathcal{H}_{i,t-1}$ denote the pre-treatment history of unit $i$, including lagged outcomes and predetermined covariates. The baseline outcome process can be written as

$$Y_{it}(0,0) = m_t(\mathcal{H}_{i,t-1}) + u_{it}, \qquad t < T_0,$$

where $m_t(\cdot)$ is an unknown forecasting function, which can be approximated using flexible prediction methods such as random forests, gradient boosting, or LASSO. MLCM estimates $m_t(\cdot)$ using pre-treatment data only, selecting the prediction algorithm and its tuning parameters through rolling panel cross-validation. The estimated model is then used to forecast the post-treatment baseline counterfactual

$$\hat{Y}_{it}(0,0) = \hat{m}_t(\mathcal{H}_{i,T_0-1}), \qquad t \geq T_0,$$

where the forecast relies only on information that is not contaminated by treatment or spillovers. Causal effects are obtained by comparing the observed outcome with the forecasted baseline outcome. For any post-treatment period $t \geq T_0$, the unit-level forecast-based gap is

$$\hat{\tau}_{it} = Y_{it} - \hat{Y}_{it}(0,0).$$

Because FBCMs do not rely on post-treatment observations of untreated units to construct counterfactuals, they remain applicable even when interference contaminates all cross-sectional comparisons. However, they identify only causal objects that can be expressed relative to $Y_{it}(0,0)$. They do not, without additional structure, recover potential outcomes under alternative nonzero exposure regimes, such as $Y_{it}(0,s)$ or $Y_{it}(1,0)$.

#### 3.2.1. FBCMs assumptions

**Assumption FBCM1 (No anticipation).** For all units and all $t < T_0$,

$$Y_{it} = Y_{it}(0,0).$$

Pre-treatment outcomes are therefore realizations of the no-treatment, no-exposure potential outcome. This is the same no-anticipation condition used for CBCMs.

**Assumption FBCM2 (Structural stability of the baseline outcome process).** In the absence of treatment and spillovers, the law of motion of the baseline potential outcome $Y_{it}(0,0)$ is stable over time. In particular, the relationship linking $Y_{it}(0,0)$ to its pre-treatment history and predetermined covariates remains valid after $T_0$.

This assumption rules out structural breaks or other shocks that would alter the baseline outcome process independently of the policy. It is the forecast-based counterpart of the comparability requirement imposed on control units in CBCMs.

**Assumption FBCM3 (Sufficient predictability).** Pre-treatment outcomes and covariates contain enough information to approximate $Y_{it}(0,0)$ with sufficiently small prediction error in the post-treatment period. Formally, for the relevant units and post-treatment periods,

$$\hat{Y}_{it}(0,0) - Y_{it}(0,0) \to 0$$

in the relevant probability or mean-square sense as the amount of usable pre-treatment information grows.

In practice, FBCM3 requires that the baseline outcome process be persistent and forecastable. The assumption is more plausible when outcomes display stable dynamics and when the pre-treatment period is long enough to validate predictive performance out of sample. It is less plausible when outcomes are dominated by large idiosyncratic shocks or when the post-treatment period is far away from the pre-treatment window.

The credibility of FBCMs therefore depends primarily on the stability and predictability of the untreated and unexposed outcome process. FBCM1 may fail when agents anticipate the policy and adjust behavior before $T_0$. FBCM2 may fail in the presence of structural breaks, regime shifts, or time-varying shocks that alter the law of motion of $Y_{it}(0,0)$ independently of the treatment. FBCM3 may fail when the available pre-treatment history is not informative enough to generate accurate counterfactual forecasts.

### 3.2.2. FBCMs identification results

We now summarize the identifying content of FBCMs under the four interference regimes introduced in Section 2.

Case (i): No spillovers.

When potential outcomes do not depend on exposure, the forecasted baseline outcome $\hat{Y}_{it}(0,0)$ directly provides the counterfactual needed to estimate the standard ATT.

**Proposition FBCM1.** Suppose outcomes do not depend on exposure, so that $Y_{it}(d,s) = Y_{it}(d,0)$ for all $s \in \mathcal{S}$. Under Assumptions FBCM1-FBCM3, FBCMs recover $Y_{it}(0,0)$ for treated units and identify

$$ATT_t = E[Y_{it}(1,0) - Y_{it}(0,0) \mid D_i = 1].$$

Case (ii.a): Spillovers affecting a subset of units (selective interference).

When exposure is observed, FBCMs can identify causal estimands defined relative to the no-treatment, no-exposure baseline. For treated units, the forecast-based gap recovers the total effect of treatment and exposure relative to $Y_{it}(0,0)$. For untreated exposed units, it recovers the spillover effect relative to the same baseline.

**Proposition FBCM2.** Suppose exposure is observed and potential outcomes are indexed by $(D_i, S_i)$. Under Assumptions FBCM1-FBCM3, FBCMs recover $Y_{it}(0,0)$ and identify, for $s \in \mathcal{S}_1$,

$$TE_t(s) = E[Y_{it}(1,s) - Y_{it}(0,0) \mid D_i = 1, S_i = s],$$

as well as

$$ATOT_t = E[Y_{it}(1, S_i) - Y_{it}(0,0) \mid D_i = 1].$$

If Assumptions FBCM1-FBCM3 also hold for untreated units, FBCMs identify, for any positive exposure level $s \in \mathcal{S}_0$,

$$SE_t(s) = E[Y_{it}(0, s) - Y_{it}(0,0) \mid D_i = 0, S_i = s],$$

and therefore

$$ASEU_t = E[Y_{it}(0, S_i) - Y_{it}(0,0) \mid D_i = 0, S_i > 0].$$

Because $ATOT_t$ and $ASEU_t$ are identified, and because untreated unexposed units have zero contribution by definition, FBCMs also identify aggregate policy effects such as $ATTE_t$ and $PAPE_t$ when the relevant population shares are observed.

By contrast, FBCMs do not generally identify the direct effect $DE_t(s)$ or the indirect effect among treated units $IE_t(s)$, because these estimands require counterfactual outcomes under nonzero exposure regimes, namely $Y_{it}(0, s)$ and $Y_{it}(1,0)$, which are not learned from pre-treatment dynamics in which treatment and exposure are inactive.

Case (ii.b): Latent selective interference.

When exposure exists but is unobserved, FBCMs can still recover baseline counterfactual outcomes from pre-treatment dynamics. Hence, they can identify aggregate effects that do not require knowing which units are exposed. However, they cannot generally recover exposure-specific or exposure-status-dependent objects.

**Proposition FBCM3.** Suppose untreated unexposed units exist but exposure status is not observed by the researcher. Under Assumptions FBCM1-FBCM3, FBCMs identify aggregate total effects on treated units,

$$ATOT_t = E[Y_{it}(1, S_i) - Y_{it}(0,0) \mid D_i = 1],$$

and the population average policy effect,

$$PAPE_t = E[Y_{it}(D_i, S_i) - Y_{it}(0,0)].$$

However, exposure-specific or exposure-status-dependent estimands such as $TE_t(s)$, $SE_t(s)$, $ASEU_t$, and $ATTE_t$ are not generally identifiable because the researcher cannot observe the exposure strata or the set of untreated exposed units.

**Case (iii): Spillovers affecting all units.**

When all units are exposed to spillovers, FBCMs remain useful because they do not require untreated and unexposed post-treatment controls. Their identifying content depends on whether exposure intensity is observed.

**Proposition FBCM4.** Suppose all units are exposed to spillovers in the post-treatment period and exposure levels are observed. Under Assumptions FBCM1-FBCM3, FBCMs recover $Y_{it}(0,0)$ and identify total effects on treated units and spillover effects on untreated units, including $TE_t(s)$, $ATOT_t$, $SE_t(s)$, $ASEU_t$, $ATTE_t$, and $PAPE_t$, whenever the corresponding conditioning groups are observed.

**Proposition FBCM5.** Suppose all units are exposed to spillovers in the post-treatment period but exposure intensity is unobserved. Under Assumptions FBCM1-FBCM3, FBCMs still identify aggregate effects defined relative to $Y_{it}(0,0)$, such as $ATOT_t$, $ASEU_t$, $ATTE_t$, and $PAPE_t$, provided the relevant treated and untreated groups are observed. They do not generally identify exposure-specific estimands such as $TE_t(s)$ or $SE_t(s)$.

As in Case (ii.a), FBCMs do not generally identify direct-effect-type estimands such as $DE_t(s)$ or indirect effects among treated units such as $IE_t(s)$ without additional structure, because these objects require counterfactual outcomes under alternative exposure regimes.

The identification proofs presented in Section 3.2.2 are reported in Appendix D.

To summarize, when spillovers are absent or sufficiently localized, identification is feasible via CBCMs under relatively standard assumptions. By contrast, when interference is widespread or its structure is unknown, CBCMs cannot identify the causal estimands of primary interest. This limitation motivates the use of alternative approaches that do not rely on the existence of uncontaminated control units, such as FBCMs. Indeed, across these settings, FBCMs can recover most of the estimands of interest under comparatively credible assumptions, particularly for short- and medium-term effects. Table 1 summarizes the scenarios discussed above and the causal estimands that CBCMs and FBCMs can identify under the assumptions described in this section.

**Table 1. Causal estimands under alternative spillover scenarios**

| Scenario | Clean controls exist? | Are Exposed units known? | Identifiable causal estimands |
|---|---|---|---|
| No spillovers | Yes | Not applicable | CBCMs: $ATT_t$<br>FBCMs: $ATT_t$ |
| Selective interference | Yes | Yes | CBCMs: $DE_t(s)$, $TE_t(s)$, $ATOT_t$, $SE_t(s)$, $ASEU_t$, $ATTE_t$, and $PAPE_t$<br>FBCMs: $TE_t(s)$, $ATOT_t$, $SE_t(s)$, $ASEU_t$, $ATTE_t$, and $PAPE_t$ |
| | Yes | No | CBCMs: none<br>FBCMs: $ATOT_t$ and $PAPE_t$ |
| Widespread spillovers: All units exposed to spillovers | No | Yes | CBCMs: $DE_t(s)$<br>FBCMs: $TE_t(s)$, $ATOT_t$, $SE_t(s)$, $ASEU_t$, $ATTE_t$, and $PAPE_t$ |
| | No | No | CBCMs: none<br>FBCMs: $ATOT_t$, $ASEU_t$, $ATTE_t$, and $PAPE_t$ |

The data-generating process introduced in the next section instantiates all spillover scenarios within a unified panel framework, allowing us to assess how identification and estimator performance evolve as interference becomes increasingly pervasive.

## 4. Simulations

### 4.1. Simulated design

This section describes the data-generating process (DGP) used to evaluate alternative estimators when the researcher either correctly observes or misspecifies the exposure mapping. The design uses the actual geography of counties in the contiguous United States and the District of Columbia.[14] In each Monte Carlo replication, 200 counties are randomly selected for direct treatment. The balanced panel contains five pre-treatment periods and one post-treatment period, with treatment beginning in period $T_0 = 6$. Let $D_i$ denote the time-invariant treatment-group indicator. The realized treatment status is therefore

$$D_{it} = D_i \, \mathbf{1}\{t \geq T_0\}, \qquad T_0 = 6.$$

[14] Researcher-defined distance mappings are constructed from centroids measured in a common projected coordinate system. County and county-equivalent boundaries are drawn from the 2023 TIGER/Line files; Alaska, Hawaii, Puerto Rico, and the other U.S. territories are excluded, and geometries are projected to EPSG:5070.

The true spillover propagation mechanism is queen contiguity: county *j* can affect county *i* when the two counties share at least one boundary point. Let $\mathcal{N}_i^Q$ denote the set of queen-contiguous neighbors of county *i*. True exposure intensity is the number of directly treated counties in this set:

$$S_i = \sum_{j \in \mathcal{N}_i^Q} D_j.$$

Untreated counties with $S_i > 0$ are exposed to spillovers, whereas untreated counties with $S_i = 0$ are unexposed and constitute clean controls under the true mapping. Directly treated counties may also be exposed when they share a border with other treated counties. The benchmark scenario, labeled "exposure mapping known," gives the researcher this true queen-contiguity mapping.

The two misspecified scenarios leave the DGP unchanged but classify exposure using distance thresholds $c \in \{30, 60\}$ km. For every county, the researcher counts treated counties whose centroids are within $c$ kilometers, excluding the county itself:

$$\mathcal{N}_i^{\text{obs}}(c) = \{j \neq i: \text{dist}(p_i, p_j) \leq c\}.$$

$$S_i^{\text{obs}}(c) = \sum_{j \in \mathcal{N}_i^{\text{obs}}(c)} D_j, \quad c \in \{30, 60\} \text{ km}.$$

A county is researcher-classified as exposed when $S_i^{\text{obs}}(c) > 0$. Outcomes are always generated by true queen exposure, but the estimators use the researcher-observed mapping to define exposure-specific target groups and apparently clean controls. Misspecification therefore produces two types of classification error: false negatives, consisting of truly exposed counties classified as unexposed, and false positives, consisting of truly unexposed counties classified as exposed.

County-specific baseline heterogeneity depends on standardized log area and standardized projected coordinates. Let $A_i$, $LON_i$, and $LAT_i$ denote these standardized characteristics. The unit effect is

$$\alpha_i = 5 + 0.55A_i - 0.30LON_i + 0.25LAT_i + u_i, \qquad u_i \sim \text{N}(0,4).$$

Directly treated counties also receive a persistent level gap of 0.5. This generates systematic pre-treatment level differences between treated and untreated counties while preserving parallel untreated-outcome trends in expectation. There is no deterministic common trend in the calibration; common time shocks are independently drawn from a normal distribution with standard deviation 1.

Two time-varying covariates, indexed by k = 1, 2, follow autoregressive processes with persistence parameter $\rho_x = 0.5$. Their initial states and innovations are independently drawn from a standard normal distribution:

$$x_{k,it} = \rho_x \, x_{k,i,t-1} + \varepsilon_{k,it}$$

Here, $X_{it} = (x_{1,i,t-1}, x_{1,i,t-2}, x_{2,i,t-1}, x_{2,i,t-2})'$ collects the four lagged covariates, with coefficient vector $\beta = (0.4, 0.2, -0.3, 0.1)'$. Let $\text{Post}_t = 1\{t \geq T_0\}$. The structural mean combines unit heterogeneity, common time shocks, the treatment-group level gap, direct treatment, spillovers, lagged covariates, and a nonlinear covariate component:

$$\mu_{it} = \alpha_i + \lambda_t + 0.5D_i + \text{direct}_{it} + \text{spillover}_{it} + \beta' X_{it} + h(X_{it}).$$

$$\text{direct}_{it} = D_i \text{Post}_t \frac{\tau}{1-\rho}, \qquad \tau = 5,$$

$$\text{spillover}_{it} = \text{Post}_t \, S_i \, \gamma_0 \, \delta, \qquad \gamma_0 = 1, \quad \delta = 3.$$

$$h(X_{it}) = 0.35({x_{1,i,t-1}}^2 - {x_{2,i,t-1}}^2) + 0.25 x_{1,i,t-1} x_{2,i,t-1}.$$

The linear lag coefficients are 0.4 and 0.2 for the first two lags of $x_1$, and −0.3 and 0.1 for the first two lags of $x_2$, respectively. Observed outcomes follow an AR(1) process with $\rho = 0.60$ and Gaussian innovations with standard deviation 1:

$$Y_{i1} = \mu_{i1} + \varepsilon_{i1}; \quad Y_{it} = \rho Y_{i,t-1} + (1-\rho)\mu_{it} + \varepsilon_{it}, \quad t \geq 2.$$

Because the direct-treatment shift in the structural mean is scaled by $1/(1-\rho)$, its realized effect in the first post-treatment period is 5. The realized spillover contribution per treated queen-contiguous neighbor is $(1-\rho)\gamma_0 \, \delta = 1.2$. Since the design contains a single post-treatment period, the true unit-level policy effect is therefore

$$\theta_i = 5D_i + 1.2S_i.$$

Simulation truths are always computed using the true queen-contiguity exposure count. The reported estimands are TE(1), the mean total effect among directly treated counties with true exposure count S = 1; ATOT, the mean total effect among all directly treated counties; SE(1), the mean spillover effect among untreated counties with true S = 1; ASEU, the mean spillover effect among truly exposed untreated counties; and PAPE, the population average policy effect.

$$\text{PAPE} = \frac{1}{N}\sum_{i=1}^{N} \theta_i = \frac{1}{N}\sum_{i=1}^{N}(5D_i + 1.2S_i).$$

Under the 30 km and 60 km scenarios, all estimators use the researcher-observed exposure count to define the TE(1), SE(1), and ASEU target groups. Estimates are evaluated against the corresponding truths defined by the queen-contiguity mapping. The design therefore captures two distinct consequences of misspecification: contamination of the apparently clean control group and changes in the composition of exposure-specific target populations. ATOT is unaffected by target-group reclassification because it averages over all directly treated counties, although its control-based estimate remains vulnerable to contaminated controls.

The control-based estimator is the doubly robust Callaway–Sant'Anna estimator (CS-DR). It uses counties classified by the researcher as untreated and unexposed as never-treated controls and adjusts for the first two lags of both time-varying covariates, standardized log area, and standardized projected coordinates. With one post-treatment period, the reported simple aggregate is the period-6 group-time effect. The control-based PAPE combines the estimated ATOT and ASEU using the researcher-defined population shares and assigns zero effect to counties classified as clean.

Forecast-based effects are estimated using two approaches. First, the MLCM predicts the untreated outcome in the post-treatment period using the two most recent outcome lags, time, four lagged covariates, and three fixed county characteristics. The model uses a random forest with 500 trees and considers three of the ten predictors as candidates at each split. Second, we use the Forecasted Average Treatment effect (FAT) estimator proposed by Botosaru et al. (2026). For each county, FAT fits a linear time trend to the five pre-treatment outcomes and extrapolates it to period 6 to estimate the outcome that would have occurred in the absence of the policy. For both methods, the unit-level effect is the observed outcome minus its forecast. Exposure-specific averages use the researcher's target classification, whereas forecast-based PAPE is the mean forecast error over the full county population and therefore does not depend on the assumed exposure mapping.

Table 2 compares the researcher's centroid distance classifications with the true spillover exposure generated by queen contiguity. Entries report the average number of counties in each classification category across 200 Monte Carlo replications. In every replication, 200 counties are directly treated and 2,909 are untreated. Figure 1 illustrates one representative replication, showing the same treatment assignment under the true exposure mapping and under the 30 km and 60 km distance-based mappings. The figure highlights correctly classified exposed and unexposed counties, as well as false negatives and false positives generated by the misspecified mappings.

**Table 2. Average numbers of counties by treatment group and exposure-classification outcome**

| Scenario | Group | Correctly exposed | False negative | False positive | Correctly unexposed |
|---|---|---|---|---|---|
| 30 km | Treated | 6.995 | 56.270 | 0.395 | 136.340 |
| 30 km | Untreated | 102.085 | 826.105 | 6.840 | 1,973.970 |
| 60 km | Treated | 49.610 | 13.655 | 13.015 | 123.720 |
| 60 km | Untreated | 722.655 | 205.535 | 189.945 | 1,790.865 |

*Notes: "Correctly exposed" and "false negative" counties are truly exposed under queen contiguity; "false positive" and "correctly unexposed" counties are truly unexposed. The researcher classifies exposure using centroid distance thresholds of 30 or 60 km.*

**Figure 1. Exposure classification under alternative researcher mappings**

Exposure mapping known
30 km
60 km
Directly treated
True exposed: correctly classified
False negative: exposed but missed
False positive: wrongly exposed
True unexposed: correctly classified

*Notes: All panels use the same simulated assignment of 200 directly treated counties. True spillover exposure is generated by queen contiguity. The first panel gives the researcher the true mapping; the remaining panels classify exposure using 30 km and 60 km distance bands between county centroids. Orange counties are truly exposed and correctly classified, blue counties are false negatives, purple counties are false positives, and light gray counties are truly unexposed and correctly classified. For visual clarity, directly treated counties are always shown in dark red regardless of their own exposure classification.*

The 30 km mapping is highly conservative. Among untreated counties, on average, it correctly identifies only 102.085 of the 928.190 counties that are truly exposed, while 826.105 are misclassified as unexposed. This corresponds to a sensitivity of approximately 11.0%. Its specificity is nevertheless very high: only 6.840 truly unexposed untreated counties are incorrectly labeled as exposed. The same pattern appears among treated counties. Consequently, the 30 km rule leaves a large number of spillover-affected counties in the apparently clean group. Expanding the distance threshold to 60 km substantially improves the detection of true exposure. For untreated counties, correctly classified exposed units rise to 722.655, producing a sensitivity of approximately 77.9%. This gain comes at the cost of more false positives: 189.945 truly unexposed untreated counties are classified as exposed. A similar trade-off is observed among treated counties, where false negatives decline sharply but false positives increase. Overall, the 30 km mapping mainly understates the reach of spillovers, whereas the 60 km mapping provides a more balanced approximation but still differs materially from the true border-based propagation mechanism.

For CBCMs, false negatives are especially consequential because truly exposed untreated counties may enter the comparison group as apparently clean controls. The 30 km mapping therefore generates substantial control-group contamination, while the 60 km mapping reduces, but does not eliminate, this problem. More generally, both false negatives and false positives alter the composition of the groups used to estimate $TE(1)$, $SE(1)$, and $ASEU$, so the resulting estimates may refer to parameters that differ from the corresponding estimands defined under the true queen-contiguity mapping. For forecast-based methods, exposure-map misspecification does not affect the construction of unit-level no-policy forecasts or the full-population $PAPE$, but it does affect the aggregation of exposure-specific effects whenever target groups are defined using the researcher's mapping.

### 4.2. Simulation results

Table 3 reports the Monte Carlo results for the county-based exposure-mapping experiment, averaged over 200 replications. The first block corresponds to the benchmark in which the researcher observes the true queen-contiguity mapping, while the 30 km and 60 km blocks leave the outcome DGP unchanged but define exposure, target groups, and apparently clean controls using the researcher's representative-point distance mapping. The first numeric column reports the average true effect across replications, and all remaining entries report Monte Carlo bias, with the standard deviation of the estimates in parentheses. The simulation code also records mean estimates, root mean squared errors, and the number of valid replications.

**Table 3. Monte Carlo bias relative to the true-mapping estimands and standard deviation under known and misspecified exposure mappings**

| | | Exposure mapping known | | | 30 km mapping | | | 60 km mapping | | |
|---|---|---|---|---|---|---|---|---|---|---|
| **Estimand** | **True value** | **CS-DR** | **MLCM** | **FAT** | **CS-DR** | **MLCM** | **FAT** | **CS-DR** | **MLCM** | **FAT** |
| **TE(1)** | 6.200 | 0.009 (0.157) | 0.097 (0.583) | 0.062 (0.679) | −0.064 (0.601) | 0.303 (0.828) | 0.277 (0.967) | −0.287 (0.196) | −0.119 (0.587) | −0.139 (0.692) |
| **ATOT** | 5.453 | 0.009 (0.114) | 0.094 (0.566) | 0.065 (0.649) | −0.405 (0.117) | 0.094 (0.566) | 0.065 (0.649) | −0.107 (0.118) | 0.094 (0.566) | 0.065 (0.649) |
| **SE(1)** | 1.200 | 0.003 (0.050) | 0.026 (0.552) | 0.060 (0.627) | −0.129 (0.164) | 0.185 (0.551) | 0.231 (0.646) | −0.314 (0.062) | −0.202 (0.549) | −0.166 (0.622) |
| **ASEU** | 1.414 | 0.001 (0.051) | 0.026 (0.551) | 0.060 (0.625) | −0.299 (0.167) | 0.011 (0.557) | 0.059 (0.650) | −0.376 (0.071) | −0.260 (0.550) | −0.220 (0.624) |
| **PAPE** | 0.772 | 0.001 (0.017) | 0.028 (0.551) | 0.058 (0.625) | −0.408 (0.011) | 0.028 (0.551) | 0.058 (0.625) | −0.124 (0.023) | 0.028 (0.551) | 0.058 (0.625) |

*Notes: The True value column reports the average causal estimand under the true queen-contiguity mapping. All other entries are Monte Carlo bias, with the standard deviation of the estimates in parentheses. In the 30 km and 60 km scenarios, exposure-specific targets and CS-DR controls are defined using the researcher's centroid distance mapping. For TE(1), SE(1), and ASEU, bias is measured relative to the intended estimand under the true mapping and therefore also captures target-group mismatch. Forecast-based PAPE is always averaged over the full population.*

When the true mapping is observed, CS-DR is essentially unbiased for every estimand and is more precise than the forecast-based alternatives. Its absolute bias never exceeds 0.009, and its standard deviations range from 0.017 for PAPE to 0.157 for TE(1). MLCM and FAT are also centered close to the truth, but their standard deviations are substantially larger. This is the setting in which CBCMs have their strongest comparative advantage: clean controls exist, are correctly identified, and can be used efficiently in cross-unit comparisons.

The 30 km rule misses most truly exposed counties and therefore leaves many spillover-affected untreated units in the apparently clean comparison group. Consistent with this contamination mechanism, CS-DR underestimates ATOT, ASEU, and PAPE, with biases of -0.405, -0.299, and -0.408, respectively. The smaller bias for TE(1) and SE(1) should not be read as evidence that the mapping is adequate: the corresponding target groups are also redefined by the observed exposure count, and the standard deviation of TE(1) rises sharply.

For MLCM and FAT, ATOT and PAPE are unchanged across mappings because the treated population and the full-population forecast average do not depend on the exposure classification. Exposure-specific estimands are different. Their aggregation uses the researcher-defined groups, so misspecification can generate bias even though the unit-level no-policy forecasts themselves are unaffected. The near-zero ASEU bias at 30 km is therefore specific to this calibration and reflects offsetting classification errors rather than general robustness of exposure-specific effects. Their invariance for ATOT and PAPE across mappings is mechanical in this design and should not be interpreted as an additional robustness result.

Expanding the radius to 60 km reduces the number of false negatives and improves the CS-DR estimates of ATOT and PAPE: their biases fall in absolute value from 0.405 to 0.107 and from 0.408 to 0.124. This improvement comes at the cost of many more false positives. As a result, the composition of the exposure-specific target groups diverges from the true queen-contiguity groups, producing sizeable negative biases for TE(1), SE(1), and ASEU across all three methods. The simulation thus reveals a non-monotonic trade-off: a broader mapping can reduce control-group contamination while simultaneously worsening the match between the estimated and intended exposure-specific estimands. Overall, mapping misspecification affects CBCMs through contamination of the comparison group and affects all methods through the redefinition of exposure-specific target populations, whereas full-population forecast-based estimands do not depend on the assumed mapping.

To complement the selective-interference analysis in Case (ii), we also consider a Case (iii) design in which spillovers reach all counties through a continuous distance-decay function. Appendix E reports additional details on the DGP used in these simulations. The results in Table 4 illustrate the implications of widespread interference. Because every untreated county is exposed to some degree, no genuinely unaffected control group exists and the CS-DR estimator cannot be implemented. By contrast, the forecast-based methods remain applicable because they construct the no-policy counterfactual from pre-treatment outcome dynamics rather than from post-treatment control units. Both MLCM and FAT produce relatively small biases for ATOT, ASEU, and PAPE, although their Monte Carlo standard deviations remain sizeable. MLCM is slightly less biased than FAT for ASEU and PAPE, while the two methods display broadly similar finite-sample performance.

**Table 4. Monte Carlo bias relative to the true-mapping estimands and standard deviation in case of widespread spillovers**

| Estimand | True value | CS-DR | MLCM | FAT |
|---|---|---|---|---|
| ATOT | 5.932 | N.A. | 0.094 (0.571) | 0.065 (0.662) |
| ASEU | 0.937 | N.A. | 0.024 (0.555) | 0.058 (0.627) |
| PAPE | 1.259 | N.A. | 0.028 (0.555) | 0.058 (0.628) |

*Notes: The True value column reports the average causal estimand under the true distance decay mapping. All other entries are Monte Carlo bias, with the standard deviation of the estimates in parentheses. Forecast-based PAPE is always averaged over the full population.*

## 5. Empirical application

We illustrate our identification framework by revisiting the natural experiment studied by Di Tella and Schargrodsky (2004). Following the terrorist attack against the A.M.I.A. Jewish center in Buenos Aires in July 1994, the federal government assigned police protection to Jewish and Muslim institutions. This generated a sudden and geographically concentrated increase in observable police presence around protected buildings. Di Tella and Schargrodsky use this shock to estimate the deterrent effect of police on car theft, comparing blocks containing protected institutions with other blocks before and after the introduction of police protection via a difference-in-differences design. The analysis uses data from three noncontiguous neighborhoods in Buenos Aires, which together account for 6.9% of the city's population. The unit of observation is the city block, with 876 blocks observed in total. Di Tella and Schargrodsky find that visible police presence sharply reduced car theft in directly protected blocks, by about 75% relative to the control group, while finding no appreciable effect one or two blocks away. The authors interpret the estimates as evidence of a strong but highly localized deterrent effect of police on crime. However, they explicitly acknowledge that unobserved displacement of car theft cannot be ruled out. Because the location choices of car thieves are unobserved, it is difficult to determine whether thefts deterred in protected blocks were displaced to other parts of the city, either inside or outside the neighborhoods included in the sample. Due to the nature of its evaluation strategy, the original study therefore provides limited evidence on the extent to which deterrence in protected blocks translated into displacement elsewhere.

This setting is particularly useful for our purposes because it naturally raises the issue of interference. Police protection may reduce crime in the protected block, but it may also affect nearby blocks. The sign of the spillover effect is theoretically ambiguous. On the one hand, police presence may generate deterrence beyond the protected block, reducing thefts in adjacent locations. On the other hand, offenders may displace criminal activity to other unprotected blocks, increasing crime just outside the protected area. Moreover, an additional reason why even blocks located farther away from protected institutions may not provide perfectly clean controls is that the redeployment of police resources may have changed the overall spatial allocation of policing.

In this section, we replicate the original study using both a more recent difference-in-differences estimator and an FBCM. As in the original paper, we use the monthly panel of blocks observed before and after the introduction of police protection. The outcome is the number of car thefts, normalized to a 30-day-equivalent monthly count. Data are available from April 1994 and the first treatment month is August 1994, the first month in which police protection was fully implemented and known to the public.[15] We focus on the first two post-treatment months, August and September.

When applying CBCMs, we must define an exposure mapping. In doing so, we follow the original paper: blocks containing protected institutions are treated units, while blocks located one block away and two blocks away from protected institutions are considered untreated but potentially exposed units. Blocks located more than two blocks away are considered the most plausible clean-control group, although their validity as truly unaffected controls is ultimately an identifying assumption for CBCMs.

We first estimate a set of difference-in-differences models using the estimator proposed by Callaway and Sant'Anna (2021). This exercise extends the original difference-in-differences logic in two directions. First, we estimate treated-block contrasts using alternative control groups: all untreated blocks, one-block controls, one- and two-block controls, and blocks more than two blocks away. In the notation of Sections 2 and 3, these contrasts recover the average total effect on treated blocks (ATOT) only when the comparison group provides a valid no-treatment/no-exposure counterfactual. Second, we estimate exposure-specific spillover effects by comparing one-block and two-block units with blocks more than two blocks away, as well as the aggregate ASEU. We then combine these estimates to obtain the PAPE. The estimates are reported in Table 5.

**Table 5. Control-based estimates of the effects of police protection on car theft**

| Estimand / comparison | August | September |
|---|---|---|
| Treated-block contrast: treated blocks vs all other blocks | -0.025 (0.038) | -0.049* (0.026) |
| Treated-block contrast: treated blocks vs one-block controls | 0.050 (0.050) | -0.008 (0.041) |
| Treated-block contrast: treated blocks vs one- and two-block controls | -0.014 (0.043) | -0.042 (0.032) |
| ATOT under clean-control assumption: treated blocks vs controls more than two blocks away | -0.035 (0.045) | -0.055** (0.026) |
| SE(1 block): one-block units vs controls more than two blocks away | -0.085** (0.039) | -0.047 (0.041) |
| SE(2 blocks): two-block units vs controls more than two blocks away | 0.023 (0.033) | 0.011 (0.028) |
| ASEU: one- and two-block units vs controls more than two blocks away | -0.022 (0.027) | -0.013 (0.027) |
| PAPE: population average policy effect | -0.011 (0.012) | -0.008 (0.012) |

*Notes: Standard errors are reported in parentheses. The outcome is the number of car thefts normalized to a 30-day-equivalent monthly count. The Callaway-Sant'Anna estimates are doubly robust difference-in-differences estimates obtained under alternative definitions of the comparison group. The treated-block rows are conventional ATT contrasts; in the notation of Sections 2 and 3, they recover ATOT only when the chosen comparison group is a valid clean-control group. The number of units in each group is as follows: 37 treated blocks, 161 blocks located one block away from a treated block, 226 blocks located two blocks away from a treated block, and 452 blocks located more than two blocks away from a treated block. The standard error of the PAPE is computed using a delta-method approximation that treats the ATOT and ASEU estimates as independent components. *, **, and *** denote statistical significance at the 10%, 5%, and 1% levels, respectively.*

The difference-in-differences estimates suggest a negative effect of police protection on car theft in the protected block, but the magnitude depends on the control group. When treated blocks are compared with all other blocks, the estimated treated-block contrast is -0.025 in August and -0.049 in September (the latter statistically significant at the 10% level). When the comparison group is restricted to blocks more than two blocks away, the estimates become slightly more negative, equal to -0.035 in August and -0.055 in September (the latter statistically significant at the 5% level). By contrast, when treated blocks are compared only with one-block controls, the estimates are not statistically significant and are even positive in one case: 0.050 in August and -0.008 in September. When two-block controls are also included, the estimates become more negative, but remain statistically insignificant. This pattern is informative. If one-block or two-block units are themselves affected by police protection, either through deterrence or displacement, using them as controls does not recover the untreated and unexposed counterfactual for treated blocks, so it delivers a biased estimate of the ATOT. The treated-block contrast is then likely contaminated by spillover effects operating within the control group.

[15] Following the original study, we exclude car thefts occurring between July 18 and July 31 and use the first 17 days of July as the final pre-treatment period.

The estimated exposure-specific spillover effects obtained with the control-based design are also unstable across distance groups. One-block units display negative effects relative to blocks more than two blocks away, equal to -0.085 at T and -0.047 at T+1 (the former statistically significant at the 5% level). Two-block units instead display small positive but not statistically significant effects. These results are consistent with a setting in which the spatial reach and even the sign of interference may vary with distance from the treated block. However, their causal interpretation relies on the assumption that blocks more than two blocks away are valid clean controls. This assumption cannot be directly verified using CBCMs: if the policy altered crime opportunities, offender mobility, or police allocation more broadly within the neighborhood, then even the "clean" comparison group may be partially contaminated.

This is precisely the setting in which CBCMs become fragile. These methods recover the no-treatment/no-exposure counterfactual by comparing treated or exposed units with observed untreated units that are assumed to be unaffected by the policy. Their validity therefore requires not only the existence of clean controls, but also information on which untreated units are exposed and which are unexposed to spillovers. In the present application, distance from the protected institution is an intuitive exposure mapping, but it may be an imperfect one. Criminal displacement and deterrence are behavioral responses that need not stop exactly after one or two blocks. As a result, the application is close to the latent or widespread interference cases discussed in Sections 2 and 3 (Cases ii.b and iii): the researcher can define plausible exposure rings, but cannot be certain that the units outside those rings are truly unaffected.

We then estimate the same policy-relevant causal objects using an FBCM: the Machine Learning Control Method (MLCM) with bagging. The model uses two lagged values of the outcome as predictors and estimates counterfactual car thefts in August and September. Unlike the control-based estimators, this approach does not construct the counterfactual for treated and exposed units from post-treatment outcomes observed in a control group. Instead, it forecasts each unit's no-treatment/no-exposure outcome using pre-treatment dynamics. This distinction is crucial in settings where most or all potential cross-sectional comparisons may themselves be affected, either directly or indirectly, by the treatment.

The MLCM estimates reported in Table 6 point to a stable negative total effect on directly protected blocks. In the notation of Sections 2 and 3, this is an estimate of ATOT rather than a direct-effect parameter $DE_t(s)$, because the forecast-based gap compares the realized outcome of treated blocks with the no-treatment/no-exposure counterfactual $Y_{it}(0,0)$. The estimated ATOT is -0.049 in August and -0.046 in September (the latter statistically significant at the 1% level). It is worth noting that these estimates are close to the Callaway and Sant'Anna treated-block contrasts obtained either using all untreated blocks as controls (row 1 of Table 5) or using only blocks more than two blocks away as controls (row 4 of Table 5). However, unlike those control-based estimates, they do not require these blocks to be valid post-treatment controls.

**Table 6. Forecast-based estimates of the effects of police protection on car theft**

| Estimand | August | September |
|---|---|---|
| ATOT: directly protected blocks | -0.049 (0.034) | -0.046*** (0.014) |
| SE(1 block): one block away from treated units | -0.035** (0.018) | -0.032 (0.023) |
| SE(2 blocks): two blocks away from treated units | 0.064*** (0.019) | 0.054*** (0.017) |
| SE(>2 blocks): more than two blocks away from treated units | 0.041*** (0.013) | 0.030** (0.012) |
| PAPE | 0.029*** (0.009) | 0.022** (0.009) |

*Notes: Standard errors are reported in parentheses. The outcome is the number of car thefts normalized to a 30-day-equivalent monthly count. The MLCM estimates are forecast-based estimates obtained using bagging with two lagged values of the outcome as predictors and 1,000 trees. The number of units in each group is as follows: 37 treated blocks, 161 blocks located one block away from a treated block, 226 blocks located two blocks away from a treated block, and 452 blocks located more than two blocks away from a treated block. *, **, and *** denote statistical significance at the 10%, 5%, and 1% levels, respectively.*

The MLCM also provides estimates of exposure-specific spillover effects by distance group. The estimated effect for one-block units is negative, equal to -0.035 in August and -0.032 in September (the former statistically significant at the 5% level), suggesting that the deterrent effect of police protection may extend slightly beyond the protected block. By contrast, the estimated effects for units two and more blocks away are positive and statistically significant at the 1% or 5% level, which is consistent with sizable and detectable displacement of car theft toward slightly more distant blocks.

Finally, we compute the Population Average Policy Effect (PAPE), which summarizes the average effect of the realized policy configuration over the entire population of blocks. The estimated PAPEs are statistically significant at the 1% or 5% level and are equal to 0.029 in August and 0.022 in September. This result highlights the difference between the local total effect on directly protected blocks and the overall population-level impact of the intervention. Police protection appears to reduce car theft in protected blocks, but the population-wide effect is positive because directly treated blocks represent a small share of all blocks and because spillover effects across untreated blocks, including units located more than two blocks away, more than offset the reduction among treated blocks. This finding is in line with the hypothetical example presented by Auerbach et al. (2024) in an experimental context.[16]

Taken together, the results illustrate the paper's main message. Control-based methods are informative when credible clean controls can be identified, but their interpretation becomes fragile when treatment may contaminate the comparison group. In this application, DiD estimates vary with the chosen controls, and the assumption that blocks more than two blocks away are unaffected is hard to defend. The forecast-based approach offers a complementary estimate of the total effect under the realized policy configuration, without relying on untreated post-treatment outcomes as clean counterfactuals. Its credibility instead rests on stable and predictable untreated outcome dynamics, an assumption that seems reasonable here given the short post-treatment horizon and the availability of monthly pre-treatment crime data.

[16] Table F1 reports the MLCM placebo estimates for the last pre-treatment period. Most are small and statistically insignificant, except for a positive estimate for blocks one block away that is marginally significant at the 10% level. Overall, the results provide little evidence of pre-treatment effects and support the credibility of the main estimates. Appendix F also provides additional details on the empirical application.

## 6. Conclusions

This paper studies the identification and estimation of causal effects in the presence of interference, a setting in which the treatment assigned to one unit may affect the outcomes of others. We develop a unified potential-outcomes framework that clarifies which causal estimands are well defined under alternative spillover regimes, depending on whether the propagation of spillovers is known or latent, and under which assumptions these estimands can be identified. In particular, we compare two broad classes of counterfactual strategies: control-based counterfactual methods (CBCMs), which rely on cross-unit comparisons, and forecast-based counterfactual methods (FBCMs), which recover untreated potential outcomes from pre-treatment dynamics.

Our theoretical results show that the viability of CBCMs hinges critically on the existence and observability of clean control units. When spillovers are absent or sufficiently localized and the exposure mapping is correctly observed, standard approaches such as DiD remain valid under a reasonable set of assumptions. However, as interference becomes more pervasive or exposure status is unobserved, the identifying variation required by cross-unit comparisons weakens or disappears. In contrast, FBCMs do not require uncontaminated controls and can recover aggregate effects defined relative to the no-treatment/no-exposure counterfactual even when exposure status is unknown. Exposure-specific spillover effects, however, remain unidentified when the relevant exposure groups cannot be observed. This identifying strength comes at a cost: FBCMs rely on the structural stability and predictability of untreated outcomes and do not, in general, allow for a clean decomposition of direct and indirect effects.

The simulation analysis provides evidence consistent with the identification results. In settings with no spillovers, or with limited and correctly observed spillovers, both CBCMs and FBCMs deliver approximately unbiased estimates of the main causal estimands of interest. As the exposure mapping becomes misspecified, control-based estimates are affected by contamination of the comparison group. Forecast-based estimates of ATOT and PAPE remain centered in this design, whereas exposure-specific averages can be biased because the researcher's mapping changes the composition of the target groups.

The empirical application revisits Di Tella and Schargrodsky's (2004) study of police protection and car theft in Buenos Aires after the 1994 terrorist attack against the A.M.I.A. Jewish center. This setting is well suited to our framework because police protection may have reduced theft in protected blocks while displacing crime or generating deterrence spillovers nearby. We compare distance-based DiD estimates with forecast-based estimates from the MLCM. The results illustrate the empirical trade-off: CBCM estimates are sensitive to how clean controls are defined, whereas FBCMs provide a complementary estimate of the total effect under the realized policy configuration.

Overall, the paper highlights a fundamental trade-off in causal inference under interference. One can rely on assumptions about the structure and reach of spillovers to preserve control-based identification, or one can shift the identifying burden toward modeling and forecasting untreated outcome dynamics. By making this trade-off explicit, we aim to contribute to a more transparent and methodologically coherent approach to policy evaluation in settings characterized by interference and spillover effects.

## References

Agrawal, A., McHale, J., & Oettl, A. (2022). How to measure the geographic spillovers of innovation? *Journal of Urban Economics, 128*, 103389.

Aguirregabiria, V., & Mira, P. (2010). Dynamic discrete choice structural models: A survey. *Journal of Econometrics, 156*(1), 38–67.

Angelucci, M., & Di Maro, V. (2016). Program evaluation and spillover effects. *Journal of Development Effectiveness, 8*(1), 22–43.

Anselin, L. (1988). *Spatial econometrics: Methods and models*. Dordrecht: Kluwer Academic Publishers.

Aronow, P. M., & Samii, C. (2017). Estimating average causal effects under general interference, with application to a social network experiment. *The Annals of Applied Statistics, 11*(4), 1912–1947.

Auerbach, E., Auerbach, J., & Tabord-Meehan, M. (2024). Discussion of "Causal inference with misspecified exposure mappings: Separating definitions and assumptions." *Biometrika*, 111(1), 21–24.

Bai, J., & Li, K. (2021). Dynamic spatial panel data models with common shocks. *Journal of Econometrics, 222*(1), 172–196.

Blundell, R., & Costa Dias, M. (2009). Alternative approaches to evaluation in empirical microeconomics. *Journal of Human Resources, 44*(3), 565–640.

Botosaru, I., Giacomini, R., & Weidner, M. (2026). Forecasted treatment effects with short panels. *arXiv preprint*, arXiv:2309.05639. Available at https://doi.org/10.48550/arXiv.2309.05639

Box, G. E. P., & Tiao, G. C. (1975). Intervention analysis with applications to economic and environmental problems. *Journal of the American Statistical Association*, 70(349), 70–79.

Bradt, J. T., & Aldy, J. E. (2025). Private benefits from public investment in climate adaptation and resilience. *NBER Working Paper* No. 33633. Available at https://www.nber.org/papers/w33633

Bramoullé, Y., Djebbari, H., & Fortin, B. (2009). Identification of peer effects through social networks. *Journal of Econometrics, 150*(1), 41–55.

Butts, K. (2023). Difference-in-differences estimation with spatial spillovers. *arXiv preprint*, arXiv:2105.03737. Available at https://doi.org/10.48550/arXiv.2105.03737

Callaway, B., & Sant'Anna, P. H. C. (2021). Difference-in-differences with multiple time periods. *Journal of Econometrics*, 225(2), 200–230.

Cerqua, A., Di Stefano, R., Letta, M., & Miccoli, S. (2021). Local mortality estimates during the COVID-19 pandemic in Italy. *Journal of Population Economics*, 34(4), 1189–1217.

Cerqua, A., Letta, M., & Menchetti, F. (2024). Causal inference and policy evaluation without a control group. *arXiv preprint*, arXiv:2312.05858. Available at https://doi.org/10.48550/arXiv.2312.05858

Cox, D. R. (1958). *Planning of experiments*. New York, NY: Wiley.

Debarsy, N., & Le Gallo, J. (2025). Spatial Spillovers: Do's and Don'ts. *Journal of Economic Surveys*, 39(5), 2152–2173.

de Paula, Á. (2017). Econometrics of network models. In B. Honoré, A. Pakes, M. Piazzesi, & L. Samuelson (Eds.), *Advances in Economics and Econometrics: Eleventh World Congress* (Vol. 1, pp. 268–323). Cambridge University Press.

Diamond, R., & McQuade, T. (2019). Who wants affordable housing in their backyard? An equilibrium analysis of low-income property development. *Journal of Political Economy*, 127(3), 1063–1117.

Di Stefano, R., & Mellace, G. (2024). The inclusive Synthetic Control Method. *arXiv preprint*, arXiv:2403.17624. Available at https://doi.org/10.48550/arXiv.2403.17624

Di Tella, R., & Schargrodsky, E. (2004). Do police reduce crime? Estimates using the allocation of police forces after a terrorist attack. *American Economic Review*, 94(1), 115–133.

Forastiere, L., Airoldi, E. M., & Mealli, F. (2021). Identification and Estimation of Treatment and Interference Effects in Observational Studies on Networks. *Journal of the American Statistical Association*, *116*(534), 901–918.

Garin, A., & Rothbaum, J. L. (2025). The long-run impacts of public industrial investment on local development and economic mobility: Evidence from World War II. *The Quarterly Journal of Economics*, 140(1), 459–520.

Gibbons, S., & Overman, H. G. (2012). Mostly pointless spatial econometrics? *Journal of Regional Science, 52*(2), 172–191.

Goldsmith-Pinkham, P., & Imbens, G. W. (2013). Social networks and the identification of peer effects. *Journal of Business & Economic Statistics, 31*(3), 253–264.

Grossi, G., Mariani, M., Mattei, A., Lattarulo, P., & Öner, Ö. (2025). Direct and spillover effects of a new tramway line on the commercial vitality of peripheral streets: A synthetic-control approach. *Journal of the Royal Statistical Society Series A: Statistics in Society, 188*(1), 223–240.

Halloran, M. E., & Struchiner, C. J. (1995). Causal inference in infectious diseases. *Epidemiology, 6*(2), 142–151.

Heckman, J. J., & Vytlacil, E. J. (2007). Econometric evaluation of social programs, Part I: Causal models, structural models and econometric policy evaluation. In J. J. Heckman & E. E. Leamer (Eds.), *Handbook of Econometrics* (Vol. 6B, pp. 4779–4874). Amsterdam: Elsevier.

Huber, M., & Steinmayr, A. (2021). A framework for separating direct and indirect effects in the presence of interference. *Journal of Applied Econometrics, 36*(6), 713–730.

Hudgens, M. G., & Halloran, M. E. (2008). Toward causal inference with interference. *Journal of the American Statistical Association*, 103(482), 832–842.

Jackson, M. O. (2010). *Social and economic networks*. Princeton, NJ: Princeton University Press.

Kelejian, H. H., & Prucha, I. R. (2010). Specification and estimation of spatial autoregressive models with autoregressive and heteroskedastic disturbances. *Journal of Econometrics, 157*(1), 53–67.

LeSage, J., & Pace, R. K. (2009). *Introduction to spatial econometrics*. Boca Raton, FL: CRC Press.

Leung, M. P. (2022). Causal inference under network interference. *Econometrica, 90*(4), 1719–1751.

Leung, M. P. (2024). Discussion of "Causal inference with misspecified exposure mappings: Separating definitions and assumptions." *Biometrika*, 111(1), 17–20.

Lewbel, A. (2019). The identification zoo: Meanings of identification in econometrics. *Journal of Economic Literature, 57*(4), 835–903.

Manski, C. F. (1993). Identification of endogenous social effects: The reflection problem. *Review of Economic Studies, 60*(3), 531–542.

Manski, C. F. (2013). Identification of treatment response with social interactions. *The Econometrics Journal, 16*(1), S1–S23.

Mealli, F., & Viviens, J. (2025). Difference-in-differences in the presence of unknown interference. *arXiv preprint*, arXiv:2512.21176. Available at https://doi.org/10.48550/arXiv.2512.21176

Menchetti, F., Cipollini, F., & Mealli, F. (2023). *Combining counterfactual outcomes and ARIMA models for policy evaluation*. *The Econometrics Journal, 26*(1), 1–24.

Ogburn, E. L., & VanderWeele, T. J. (2017). Causal diagrams for interference. *Statistical Science, 32*(1), 62–80.

Rubin, D. B. (1974). Estimating causal effects of treatments in randomized and nonrandomized studies. *Journal of Educational Psychology, 66*(5), 688–701.

Rubin, D. B. (1980). Comment on "Randomization analysis of experimental data in the Fisher randomization test" by D. Basu. *Journal of the American Statistical Association, 75*(371), 591–593.

Sävje, F. (2024). Causal inference with misspecified exposure mappings: Separating definitions and assumptions. *Biometrika*, 111(1), 1–15.

Sävje, F., Aronow, P. M., & Hudgens, M. G. (2021). Average treatment effects in the presence of unknown interference. *The Annals of Statistics, 49*(2), 673–701.

Shi, W., & Lee, L. F. (2018). Spatial dynamic panel data models with interactive fixed effects. *Journal of Econometrics, 207*(2), 313–339.

Sobel, M. E. (2006). What do randomized studies of housing mobility demonstrate? *Journal of the American Statistical Association, 101*(476), 1398–1407.

Tchetgen Tchetgen, E. J., & VanderWeele, T. J. (2012). On causal inference in the presence of interference. *Statistical Methods in Medical Research, 21*(1), 55–75.

Viviano, D. (2025). Policy targeting under network interference. *The Review of Economic Studies*, 92(2), 1257–1292.

Xu, R. (2025). Difference-in-Differences with Interference. *arXiv preprint*, arXiv:2306.12003. Available at https://doi.org/10.48550/arXiv.2306.12003

## Appendix A. Proof of Proposition 1

**Proof of Proposition 1.** Fix a post-treatment period $t \geq T_0$ and an exposure level $s \in \mathcal{S}_1$. Recall that the total effect at exposure level $s$ for treated units is defined as

$$TE_t(s) = \mathbb{E}[Y_{it}(1, s) - Y_{it}(0,0) \mid D_i = 1, S_i = s].$$

Add and subtract $Y_{it}(1,0)$ inside the conditional expectation:

$$TE_t(s) = \mathbb{E}[Y_{it}(1, s) - Y_{it}(1,0) + Y_{it}(1,0) - Y_{it}(0,0) \mid D_i = 1, S_i = s].$$

By linearity of conditional expectation,

$$TE_t(s) = \mathbb{E}[Y_{it}(1,0) - Y_{it}(0,0) \mid D_i = 1, S_i = s] + \mathbb{E}[Y_{it}(1, s) - Y_{it}(1,0) \mid D_i = 1, S_i = s].$$

Using the definitions of $DE_t^0(s)$ and $IE_t(s)$, this implies

$$TE_t(s) = DE_t^0(s) + IE_t(s),$$

which proves the result. ▫

## Appendix B. Difference-in-differences with exposure rings

This appendix presents a compact difference-in-differences specification with exposure rings. The goal is to estimate the effect on directly treated units and the spillover effects on untreated units located at different distances from treatment.

Let the post-treatment indicator be

$$P_t = \mathbf{1}\{t \geq T_0\}.$$

For each untreated unit, let $d_i$ denote the distance to the nearest treated unit. Choose distance cutoffs

$$0 = \rho_0 < \rho_1 < \cdots < \rho_R,$$

and define the mutually exclusive ring indicators

$$R_{ir} = \mathbf{1}\{D_i = 0,\ \rho_{r-1} < d_i \leq \rho_r\}, \qquad r = 1, \ldots, R.$$

Untreated units with $D_i = 0$ and $d_i > \rho_R$ form the omitted reference group. Directly treated units are kept separate from the exposure rings.

The baseline difference-in-differences specification is

$$Y_{it} = \alpha_i + \lambda_t + \beta(D_i \times P_t) + \sum_{r=1}^{R} \delta_r \,(R_{ir} \times P_t) + \varepsilon_{it}.$$

Here $\alpha_i$ are unit fixed effects and $\lambda_t$ are time fixed effects. The coefficient $\beta$ can be interpreted as a pooled total effect on directly treated units relative to the omitted clean-control group. The coefficients $\delta_r$ capture the post-treatment change for untreated units in ring $r$ relative to the same reference group. The sequence $\delta_1, \ldots, \delta_R$ describes how spillover effects vary with distance from treated units.

In the notation of Section 2, the ring indicators are a discretized version of exposure $S_i$. The omitted reference group corresponds to $S_i = 0$, while the rings correspond to positive exposure categories among untreated units.

With several pre- and post-treatment periods, the same logic can be written as

$$Y_{it} = \alpha_i + \lambda_t + \sum_{\ell \neq -1} \beta_\ell \,[D_i \times \mathbf{1}\{t - T_0 = \ell\}] + \sum_{r=1}^{R} \sum_{\ell \neq -1} \delta_{r\ell} \,[R_{ir} \times \mathbf{1}\{t - T_0 = \ell\}] + \varepsilon_{it}.$$

This corresponds to the event-study version of this design. The omitted event time is $\ell = -1$. The coefficients $\beta_\ell$ trace the dynamic path for directly treated units, while $\delta_{r\ell}$ traces the dynamic spillover path for untreated units in ring $r$.

Rings are generally defined before treatment using geography, network distance, travel time, commuting links, or another theoretically motivated measure of exposure. The outer reference group should be plausibly far enough from the treated units to be unaffected by the intervention. Standard errors should be clustered at the relevant assignment or geographical level; spatially robust standard errors may be preferable when residual spatial correlation is likely.

This specification keeps potentially exposed untreated units out of the clean-control group and allows the researcher to assess whether spillover effects decay, disappear, or change sign with distance from treatment.

## Appendix C. Identification proofs for CBCMs

This appendix provides the proofs of the identification results reported in Section 3.1.3. Let $t \geq T_0$ denote a post-treatment period and let $T_0 - 1$ denote the last pre-treatment period. Under no anticipation, pre-treatment outcomes satisfy $Y_{it} = Y_{it}(0,0)$. To keep the notation compact, define the untreated and unexposed potential outcome as

$$Y_{it}^0 \equiv Y_{it}(0,0).$$

Clean controls are untreated and unexposed units, i.e. units with $D_i = 0$ and $S_i = 0$.

Throughout, we maintain the consistency relation, $Y_{it} = Y_{it}(D_i, S_{it})$. Since $t \geq T_0$, $S_{it} = S_i$, while no anticipation implies $Y_{i,T_0-1} = Y_{i,T_0-1}(0,0)$. We invoke these relations below without repeating them.

Case (i): No spillovers.

**Proof of Proposition CBCM1**

Under no spillovers, potential outcomes do not depend on exposure, so $Y_{it}(d,s) = Y_{it}(d,0) = Y_{it}(d)$ for all $s \in \mathcal{S}$. Consider the standard DiD estimand:

$$\Delta_t^{\text{DID}} = \mathbb{E}\left[Y_{it} - Y_{i,T_0-1} \mid D_i = 1\right] - \mathbb{E}\left[Y_{it} - Y_{i,T_0-1} \mid D_i = 0\right].$$

By no anticipation, and no spillovers,

$$\Delta_t^{\text{DID}} = \mathbb{E}\left[Y_{it}(1,0) - Y_{i,T_0-1}^0 \mid D_i = 1\right] - \mathbb{E}\left[Y_{it}^0 - Y_{i,T_0-1}^0 \mid D_i = 0\right].$$

Add and subtract $\mathbb{E}\left[Y_{it}^0 - Y_{i,T_0-1}^0 \mid D_i = 1\right]$. Then

$$\Delta_t^{\text{DID}} = \text{ATT}_t + B_t,$$

where

$$\text{ATT}_t = \mathbb{E}[Y_{it}(1,0) - Y_{it}^0 \mid D_i = 1]$$

and

$$B_t = \mathbb{E}\left[Y_{it}^0 - Y_{i,T_0-1}^0 \mid D_i = 1\right] - \mathbb{E}\left[Y_{it}^0 - Y_{i,T_0-1}^0 \mid D_i = 0\right].$$

In Case (i), all untreated units are valid controls because their outcomes are unaffected by exposure. Under assumption CBCM3(a), $B_t = 0$. Therefore, $\Delta_t^{\text{DID}} = \text{ATT}_t$. This proves the result.

Case (ii.a): Spillovers affecting a subset of units (selective interference).

**Proof of Proposition CBCM2**

Fix $s \in \mathcal{S}_1$. Consider the DiD estimand comparing treated units with exposure $s$ to untreated and unexposed units:

$$\Delta_t^{\text{TE}}(s) = \mathbb{E}\left[Y_{it} - Y_{i,T_0-1} \mid D_i = 1, S_i = s\right] - \mathbb{E}\left[Y_{it} - Y_{i,T_0-1} \mid D_i = 0, S_i = 0\right].$$

By no anticipation, and the existence of clean controls,

$$\Delta_t^{\mathrm{TE}}(s) = \mathbb{E}\left[Y_{it}(1,s) - Y_{i,T_0-1}^0 \mid D_i = 1, S_i = s\right] - \mathbb{E}\left[Y_{it}^0 - Y_{i,T_0-1}^0 \mid D_i = 0, S_i = 0\right].$$

Add and subtract $\mathbb{E}\left[Y_{it}^0 - Y_{i,T_0-1}^0 \mid D_i = 1, S_i = s\right]$. Then

$$\Delta_t^{\mathrm{TE}}(s) = \mathrm{TE}_t(s) + B_t^{\mathrm{TE}}(s),$$

where

$$\mathrm{TE}_t(s) = \mathbb{E}[Y_{it}(1,s) - Y_{it}^0 \mid D_i = 1, S_i = s]$$

and

$$B_t^{\mathrm{TE}}(s) = \mathbb{E}\left[Y_{it}^0 - Y_{i,T_0-1}^0 \mid D_i = 1, S_i = s\right] - \mathbb{E}\left[Y_{it}^0 - Y_{i,T_0-1}^0 \mid D_i = 0, S_i = 0\right].$$

Assumption CBCM3(a) implies $B_t^{\mathrm{TE}}(s) = 0$. Therefore, $\Delta_t^{\mathrm{TE}}(s) = \mathrm{TE}_t(s)$. This proves the result.

**Proof of Corollary CBCM1**

Set $s = 0$ in Proposition CBCM2. For treated units with zero exposure,

$$\mathrm{TE}_t(0) = \mathbb{E}[Y_{it}(1,0) - Y_{it}^0 \mid D_i = 1, S_i = 0].$$

This coincides with the direct effect for treated units with zero exposure, since own treatment changes from $0$ to $1$ while exposure is fixed at $0$. Hence, under the assumptions of Proposition CBCM2, DiD comparing treated unexposed units to untreated unexposed units identifies $\mathrm{TE}_t(0) = \mathrm{DE}_t(0)$. This proves the corollary.

**Proof of Proposition CBCM3**

Consider the DiD estimand comparing all treated units to untreated and unexposed units:

$$\Delta_t^{\mathrm{ATOT}} = \mathbb{E}\left[Y_{it} - Y_{i,T_0-1} \mid D_i = 1\right] - \mathbb{E}\left[Y_{it} - Y_{i,T_0-1} \mid D_i = 0, S_i = 0\right].$$

By no anticipation, and the existence of clean controls,

$$\Delta_t^{\mathrm{ATOT}} = \mathbb{E}\left[Y_{it}(1,S_i) - Y_{i,T_0-1}^0 \mid D_i = 1\right] - \mathbb{E}\left[Y_{it}^0 - Y_{i,T_0-1}^0 \mid D_i = 0, S_i = 0\right].$$

Add and subtract $\mathbb{E}\left[Y_{it}^0 - Y_{i,T_0-1}^0 \mid D_i = 1\right]$. Then

$$\Delta_t^{\mathrm{ATOT}} = \mathrm{ATOT}_t + B_t^{\mathrm{ATOT}},$$

where

$$\mathrm{ATOT}_t = \mathbb{E}[Y_{it}(1,S_i) - Y_{it}^0 \mid D_i = 1]$$

and

$$B_t^{\mathrm{ATOT}} = \mathbb{E}\left[Y_{it}^0 - Y_{i,T_0-1}^0 \mid D_i = 1\right] - \mathbb{E}\left[Y_{it}^0 - Y_{i,T_0-1}^0 \mid D_i = 0, S_i = 0\right].$$

Assumption CBCM3(a), applied within exposure strata and averaged over the realized exposure distribution among treated units, implies $B_t^{\mathrm{ATOT}} = 0$. Therefore, $\Delta_t^{\mathrm{ATOT}} = \mathrm{ATOT}_t$. This proves the result.

**Proof of Proposition CBCM4**

Fix $s \in \mathcal{S}_0$ with $s > 0$. Consider the DiD estimand comparing untreated units with exposure $s$ to untreated and unexposed units:

$$\Delta_t^{\mathrm{SE}}(s) = \mathbb{E}\left[Y_{it} - Y_{i,T_0-1} \mid D_i = 0, S_i = s\right] - \mathbb{E}\left[Y_{it} - Y_{i,T_0-1} \mid D_i = 0, S_i = 0\right].$$

By no anticipation, and the existence of clean controls,

$$\Delta_t^{\mathrm{SE}}(s) = \mathbb{E}\left[Y_{it}(0,s) - Y_{i,T_0-1}^0 \mid D_i = 0, S_i = s\right] - \mathbb{E}\left[Y_{it}^0 - Y_{i,T_0-1}^0 \mid D_i = 0, S_i = 0\right].$$

Add and subtract $\mathbb{E}\left[Y_{it}^0 - Y_{i,T_0-1}^0 \mid D_i = 0, S_i = s\right]$. Then

$$\Delta_t^{\mathrm{SE}}(s) = \mathrm{SE}_t(s) + B_t^{\mathrm{SE}}(s),$$

where

$$\mathrm{SE}_t(s) = \mathbb{E}[Y_{it}(0,s) - Y_{it}^0 \mid D_i = 0, S_i = s]$$

and

$$B_t^{\mathrm{SE}}(s) = \mathbb{E}\left[Y_{it}^0 - Y_{i,T_0-1}^0 \mid D_i = 0, S_i = s\right] - \mathbb{E}\left[Y_{it}^0 - Y_{i,T_0-1}^0 \mid D_i = 0, S_i = 0\right].$$

Assumption CBCM3(b) implies $B_t^{\mathrm{SE}}(s) = 0$. Therefore, $\Delta_t^{\mathrm{SE}}(s) = \mathrm{SE}_t(s)$. This proves the result.

**Proof of Proposition CBCM5**

By Proposition CBCM4, $\mathrm{SE}_t(s)$ is identified for each relevant positive exposure level $s \in \mathcal{S}_0$ with $s > 0$. The Average Spillover Effect on the Exposed Untreated is

$$\mathrm{ASEU}_t = \mathbb{E}[Y_{it}(0,S_i) - Y_{it}^0 \mid D_i = 0, S_i > 0].$$

Equivalently,

$$\mathrm{ASEU}_t = \mathbb{E}[\mathrm{SE}_t(S_i) \mid D_i = 0, S_i > 0].$$

Since the exposure distribution among untreated exposed units is observed under Case (ii.a), averaging the identified exposure-specific effects $\mathrm{SE}_t(s)$ over this distribution identifies $\mathrm{ASEU}_t$. This proves the result.

**Proof of Corollary CBCM2**

Let

$$\pi_T = \Pr(D_i = 1), \qquad \pi_U = \Pr(D_i = 0, S_i > 0), \qquad \pi_C = \Pr(D_i = 0, S_i = 0).$$

Under selective interference with observed exposure, the population is partitioned into treated units, untreated exposed units, and untreated unexposed units. The average policy effect among directly or indirectly exposed units is

$$\mathrm{ATTE}_t = \frac{\pi_T \mathrm{ATOT}_t + \pi_U \mathrm{ASEU}_t}{\pi_T + \pi_U}.$$

The Population Average Policy Effect is

$$\text{PAPE}_t = \pi_T \text{ATOT}_t + \pi_U \text{ASEU}_t + \pi_C \cdot 0.$$

The last term is zero because untreated and unexposed units experience neither direct treatment nor exposure. If $\text{ATOT}_t$, $\text{ASEU}_t$, and the relevant population shares are identified, both $\text{ATTE}_t$ and $\text{PAPE}_t$ are identified by substitution. This proves the corollary.

Case (ii.b): Latent selective interference.

**Proof of Proposition CBCM6**

Under latent selective interference, untreated and unexposed units exist in the population, but exposure status is not observed by the researcher. Hence, the researcher cannot distinguish untreated unexposed units $(D_i = 0, S_i = 0)$ from untreated exposed units $(D_i = 0, S_i > 0)$. The clean-control group required to recover $Y_{it}^0$ therefore cannot be isolated in the observed data.

The observed untreated group is a mixture of units whose post-treatment outcomes are $Y_{it}^0$ and units whose post-treatment outcomes are $Y_{it}(0, S_i)$ with $S_i > 0$. Without observing exposure status, different assignments of units to exposed and unexposed regimes can generate the same observed distribution of outcomes while implying different values of $\text{TE}_t(s)$, $\text{ATOT}_t$, $\text{SE}_t(s)$, $\text{ASEU}_t$, $\text{ATTE}_t$, and $\text{PAPE}_t$. Therefore, these estimands are not generally point-identified by CBCMs. This proves the result.

Case (iii): Spillovers affecting all units.

**Proof of Proposition CBCM7**

Suppose that all units are exposed to spillovers in the post-treatment period, so that no unit satisfies $D_i = 0$ and $S_i = 0$. Then no untreated and unexposed observations are available within the observed population. For untreated units, the observed post-treatment outcome is

$$Y_{it} = Y_{it}(0, S_i), \qquad S_i > 0,$$

not $Y_{it}^0$. Hence, untreated outcomes are contaminated by spillovers and cannot be used as observations of the no-treatment/no-exposure potential outcome.

All estimands listed in Proposition CBCM7 are defined relative to $Y_{it}^0$, either directly or through averages involving this baseline counterfactual. Since CBCMs recover counterfactuals from observed cross-unit comparisons and no unit reveals $Y_{it}^0$ in the post-treatment period, CBCMs cannot generally point-identify $\text{TE}_t(s)$, $\text{ATOT}_t$, $\text{SE}_t(s)$, $\text{ASEU}_t$, $\text{ATTE}_t$, or $\text{PAPE}_t$. This proves the result.

**Proof of Proposition CBCM8**

Fix an exposure level $s$ for which there is common support between treated and untreated units. Consider the DiD estimand comparing treated and untreated units with the same exposure level:

$$\Delta_t^{\text{DE}}(s) = \mathbb{E}\left[Y_{it} - Y_{i,T_0-1} \mid D_i = 1, S_i = s\right] - \mathbb{E}\left[Y_{it} - Y_{i,T_0-1} \mid D_i = 0, S_i = s\right].$$

By no anticipation,

$$\Delta_t^{\mathrm{DE}}(s) = \mathbb{E}\left[Y_{it}(1,s) - Y_{i,T_0-1}^0 \mid D_i = 1, S_i = s\right] - \mathbb{E}\left[Y_{it}(0,s) - Y_{i,T_0-1}^0 \mid D_i = 0, S_i = s\right].$$

Add and subtract $\mathbb{E}\left[Y_{it}(0,s) - Y_{i,T_0-1}^0 \mid D_i = 1, S_i = s\right]$. Then

$$\Delta_t^{\mathrm{DE}}(s) = \mathrm{DE}_t(s) + B_t^{\mathrm{DE}}(s),$$

where

$$\mathrm{DE}_t(s) = \mathbb{E}[Y_{it}(1,s) - Y_{it}(0,s) \mid D_i = 1, S_i = s]$$

and

$$B_t^{\mathrm{DE}}(s) = \mathbb{E}\left[Y_{it}(0,s) - Y_{i,T_0-1}^0 \mid D_i = 1, S_i = s\right] - \mathbb{E}\left[Y_{it}(0,s) - Y_{i,T_0-1}^0 \mid D_i = 0, S_i = s\right].$$

Assumption CBCM3(c) sets $B_t^{\mathrm{DE}}(s) = 0$. Therefore, $\Delta_t^{\mathrm{DE}}(s) = \mathrm{DE}_t(s)$. This proves the result.

## Appendix D. Identification proofs for FBCMs

This appendix reports the identification proofs for the FBCMs discussed in Section 3.2. Throughout, let $\hat{Y}_{it}(0,0)$ denote the forecast-based counterfactual outcome constructed using only pre-treatment data. Under Assumptions FBCM1-FBCM3, the forecast recovers the baseline potential outcome for the relevant units and post-treatment periods, in the sense that

$$\hat{Y}_{it}(0,0) \to Y_{it}(0,0).$$

The corresponding forecast-based gap is

$$\hat{\Delta}_{it}^{F} = Y_{it} - \hat{Y}_{it}(0,0).$$

Case (i): No spillovers.

**Proof of Proposition FBCM1.** Suppose outcomes do not depend on exposure, so that

$$Y_{it}(d,s) = Y_{it}(d,0) = Y_{it}(d) \qquad \text{for all } s \in \mathcal{S}.$$

For treated units in the post-treatment period, consistency implies

$$Y_{it} = Y_{it}(1,0) \qquad \text{for } D_i = 1,\ t \geq T_0.$$

By Assumption FBCM1, pre-treatment outcomes are realizations of the baseline process $Y_{it}(0,0)$. By Assumptions FBCM2 and FBCM3, this process is stable and sufficiently predictable, so that the forecast-based counterfactual recovers

$$\hat{Y}_{it}(0,0) \to Y_{it}(0,0).$$

Therefore, for treated units,

$$\hat{\Delta}_{it}^{F} = Y_{it} - \hat{Y}_{it}(0,0) \to Y_{it}(1,0) - Y_{it}(0,0).$$

Taking expectations conditional on $D_i = 1$ yields

$$E\left[\hat{\Delta}_{it}^{F} \mid D_i = 1\right] \to E\{Y_{it}(1,0) - Y_{it}(0,0) \mid D_i = 1\} = ATT_t.$$

Hence, FBCMs identify $ATT_t$. ▫

Case (ii.a): Selective interference

**Proof of Proposition FBCM2.** Suppose exposure is observed and potential outcomes are indexed by $(d,s)$. For treated units with exposure level $s \in \mathcal{S}_1$, consistency implies

$$Y_{it} = Y_{it}(1,s) \qquad \text{for } D_i = 1,\ S_i = s,\ t \geq T_0.$$

As before, Assumptions FBCM1-FBCM3 imply that the forecast-based counterfactual recovers the no-treatment, no-exposure baseline,

$$\hat{Y}_{it}(0,0) \to Y_{it}(0,0).$$

Therefore,

$$\hat{\Delta}_{it}^F = Y_{it} - \hat{Y}_{it}(0,0) \rightarrow Y_{it}(1,s) - Y_{it}(0,0)$$

for treated units with $S_i = s$. Taking expectations conditional on $D_i = 1$ and $S_i = s$ gives

$$E[\hat{\Delta}_{it}^F \mid D_i = 1, S_i = s] \rightarrow E\{Y_{it}(1,s) - Y_{it}(0,0) \mid D_i = 1, S_i = s\} = TE_t(s).$$

Thus, FBCMs identify $TE_t(s)$ for all observed exposure levels among treated units. Averaging over the realized exposure distribution among treated units yields

$$E[\hat{\Delta}_{it}^F \mid D_i = 1] \rightarrow E\{Y_{it}(1,S_i) - Y_{it}(0,0) \mid D_i = 1\} = ATOT_t.$$

For untreated exposed units, suppose that Assumptions FBCM1-FBCM3 also hold for untreated units. For $D_i = 0$ and $S_i = s > 0$, consistency implies

$$Y_{it} = Y_{it}(0,s).$$

Since $\hat{Y}_{it}(0,0) \rightarrow Y_{it}(0,0)$, it follows that

$$E[\hat{\Delta}_{it}^F \mid D_i = 0, S_i = s] \rightarrow E\{Y_{it}(0,s) - Y_{it}(0,0) \mid D_i = 0, S_i = s\} = SE_t(s).$$

Averaging over the realized distribution of positive exposure among untreated units gives

$$E[\hat{\Delta}_{it}^F \mid D_i = 0, S_i > 0] \rightarrow E\{Y_{it}(0,S_i) - Y_{it}(0,0) \mid D_i = 0, S_i > 0\} = ASEU_t.$$

Finally, because $ATOT_t$ and $ASEU_t$ are identified, and because untreated unexposed units have zero contribution by definition, the aggregate estimands $ATTE_t$ and $PAPE_t$ are identified whenever the relevant population shares are observed. In particular,

$$PAPE_t = Pr(D_i = 1)ATOT_t + Pr(D_i = 0, S_i > 0)ASEU_t,$$

under selective interference, since the contribution of untreated unexposed units is zero. ▫

Case (ii.b): Latent selective interference.

**Proof of Proposition FBCM3.** Suppose untreated unexposed units exist but exposure status is not observed by the researcher. FBCMs do not require identifying untreated unexposed units to recover the baseline counterfactual. By Assumptions FBCM1-FBCM3,

$$\hat{Y}_{it}(0,0) \rightarrow Y_{it}(0,0)$$

for the relevant units. For treated units, consistency implies

$$Y_{it} = Y_{it}(1,S_i),$$

where $S_i$ is the realized but unobserved exposure level. Hence,

$$\hat{\Delta}_{it}^F = Y_{it} - \hat{Y}_{it}(0,0) \rightarrow Y_{it}(1,S_i) - Y_{it}(0,0).$$

Taking expectations conditional on $D_i = 1$ yields

$$E[\hat{\Delta}_{it}^F \mid D_i = 1] \rightarrow E\{Y_{it}(1,S_i) - Y_{it}(0,0) \mid D_i = 1\} = ATOT_t.$$

Thus, FBCMs identify aggregate total effects on treated units even when exposure status is latent.

The population average policy effect is also identified because it is defined as the average gap between the realized observed outcome and the no-treatment, no-exposure baseline over the whole population:

$$PAPE_t = E\{Y_{it}(D_i, S_i) - Y_{it}(0,0)\}.$$

Since $Y_{it} = Y_{it}(D_i, S_i)$ in the post-treatment period and $\hat{Y}_{it}(0,0) \to Y_{it}(0,0)$,

$$E[\hat{\Delta}_{it}^F] \to E\{Y_{it}(D_i, S_i) - Y_{it}(0,0)\} = PAPE_t.$$

However, exposure-specific or exposure-status-dependent estimands require observing the exposure strata or the set of exposed untreated units. Since exposure status is unobserved, the researcher cannot condition on events such as $S_i = s$ or $S_i > 0$. Therefore, $TE_t(s)$, $SE_t(s)$, $ASEU_t$, and $ATTE_t$ are not generally identifiable under latent selective interference. ▫

Case (iii): Spillovers affecting all units.

**Proof of Proposition FBCM4.** Suppose all units are exposed to spillovers in the post-treatment period and exposure levels are observed. Under Assumptions FBCM1-FBCM3, FBCMs recover the no-treatment, no-exposure baseline,

$$\hat{Y}_{it}(0,0) \to Y_{it}(0,0),$$

for the relevant units and post-treatment periods.

For treated units with exposure level $s$, consistency implies

$$Y_{it} = Y_{it}(1, s) \qquad \text{for } D_i = 1,\ S_i = s.$$

Therefore,

$$E[\hat{\Delta}_{it}^F \mid D_i = 1, S_i = s] \to E\{Y_{it}(1, s) - Y_{it}(0,0) \mid D_i = 1, S_i = s\} = TE_t(s).$$

Averaging over the realized exposure distribution among treated units identifies $ATOT_t$.

For untreated units with exposure level $s$, consistency implies

$$Y_{it} = Y_{it}(0, s) \qquad \text{for } D_i = 0,\ S_i = s.$$

Hence,

$$E[\hat{\Delta}_{it}^F \mid D_i = 0, S_i = s] \to E\{Y_{it}(0, s) - Y_{it}(0,0) \mid D_i = 0, S_i = s\} = SE_t(s).$$

Averaging over untreated exposed units identifies $ASEU_t$. Since in this case all untreated units are exposed, the conditioning event $D_i = 0, S_i > 0$ coincides with $D_i = 0$. Finally, because $ATOT_t$ and $ASEU_t$ are identified and the relevant population shares are observed, the aggregate estimands $ATTE_t$ and $PAPE_t$ are also identified. ▫

**Proof of Proposition FBCM5.** Suppose all units are exposed to spillovers in the post-treatment period but exposure intensity is unobserved. By Assumptions FBCM1-FBCM3,

$$\hat{Y}_{it}(0,0) \to Y_{it}(0,0).$$

For treated units, the observed post-treatment outcome is $Y_{it} = Y_{it}(1, S_i)$, where $S_i > 0$ is realized but unobserved. Thus,

$$E\left[\hat{\Delta}_{it}^{F} \mid D_i = 1\right] \to E\{Y_{it}(1, S_i) - Y_{it}(0,0) \mid D_i = 1\} = ATOT_t.$$

For untreated units, universal exposure implies $S_i > 0$ for all $D_i = 0$. Hence,

$$E\left[\hat{\Delta}_{it}^{F} \mid D_i = 0\right] \to E\{Y_{it}(0, S_i) - Y_{it}(0,0) \mid D_i = 0, S_i > 0\} = ASEU_t.$$

Because treated and untreated groups are observed, and because all untreated units are exposed, the relevant aggregate effects defined relative to $Y_{it}(0,0)$, including $ATOT_t$, $ASEU_t$, and $PAPE_t$, are identified. Since all units are exposed in this case, $ATTE_t$ coincides with $PAPE_t$.

However, exposure-specific estimands require conditioning on the unobserved exposure level $S_i = s$. Since exposure intensity is not observed, quantities such as $TE_t(s)$ and $SE_t(s)$ cannot generally be recovered. Moreover, direct-effect-type estimands such as $DE_t(s)$ and indirect effects among treated units such as $IE_t(s)$ require counterfactual outcomes under alternative exposure regimes, such as $Y_{it}(0, s)$ and $Y_{it}(1,0)$, which are not learned from pre-treatment dynamics without additional structure. ▫

## Appendix E. Additional details on the simulations

To represent Case (iii), in which spillovers affect all units, we consider a second simulation design based on the actual geography of counties in the contiguous United States and the District of Columbia. In each of 200 Monte Carlo replications, 200 counties are randomly assigned to direct treatment. The balanced panel contains five pre-treatment periods and one post-treatment period, with treatment beginning in period 6. The baseline outcome process is kept comparable to that used in the selective-interference simulations, but queen-contiguity exposure is replaced by a continuous distance-decay mechanism.

Let $d_{ij}$ denote the distance in kilometres between the interior representative points of counties $i$ and $j$. The spatial weight connecting the two counties is defined as

$$w_{ij} = \exp\left(-\frac{d_{ij}}{60}\right), \qquad w_{ii} = 0,$$

and the true exposure score of county $i$ is

$$S_i = \sum_{j \neq i} w_{ij} D_j.$$

Spillovers therefore decline smoothly with distance but never become exactly zero. Since the exponential weights are strictly positive, every county receives some exposure from the treated counties, although the resulting effect may be extremely small for geographically distant units. Both untreated and treated counties can experience spillovers, while the zero diagonal prevents a treated county from generating a spillover to itself.

County-specific baseline heterogeneity depends on standardized log area and standardized projected coordinates. Let $\tilde{A}_i$, $\tilde{X}_i$, and $\tilde{Y}_i$ denote these characteristics. The unit effect is

$$\alpha_i = 5 + 0.55\tilde{A}_i - 0.30\tilde{X}_i + 0.25\tilde{Y}_i + u_i, \qquad u_i \sim \mathcal{N}(0, 2^2).$$

Directly treated counties also receive a persistent level gap of 0.5. This generates systematic pre-treatment level differences between treated and untreated counties while preserving parallel untreated-outcome trends in expectation. There is no deterministic common trend, and common time shocks are independently drawn from a normal distribution with standard deviation 1.

Two time-varying covariates, indexed by $k = 1,2$, follow autoregressive processes,

$$x_{kit} = 0.5x_{ki,t-1} + \nu_{kit}, \qquad \nu_{kit} \sim \mathcal{N}(0,1).$$

The structural mean includes the first two lags of each covariate, with coefficient vector

$$\boldsymbol{\beta} = (0.4,\ 0.2,\ -0.3,\ 0.1)',$$

and the nonlinear component

$$g_{it} = 0.35\left(x_{1,i,t-1}^2 - x_{2,i,t-1}^2\right) + 0.25x_{1,i,t-1}x_{2,i,t-1}.$$

Let $P_t = \mathbb{1}\{t \geq 6\}$. The structural mean is

$$\mu_{it} = \alpha_i + \lambda_t + 0.5D_i + 12.5D_iP_t + 3S_iP_t + \boldsymbol{\beta}'\mathbf{X}_{it} + g_{it},$$

where $\lambda_t$ is the common time shock and $\mathbf{X}_{it}$ collects the four lagged covariates. Observed outcomes follow the autoregressive process

$$Y_{it} = 0.6Y_{i,t-1} + 0.4\mu_{it} + \varepsilon_{it}, \qquad \varepsilon_{it} \sim \mathcal{N}(0,1).$$

The treatment component entering the structural mean is scaled so that the realized direct effect in the first post-treatment period is 5. The realized spillover contribution is $0.4 \times 3S_i = 1.2S_i$. Consequently, the true unit-level policy effect in period 6 is

$$\theta_i = 5D_i + 1.2S_i.$$

The simulation reports the Average Total Effect on the Treated (ATOT), the Average Spillover Effect on Exposed Untreated units (ASEU), and the Population Average Policy Effect (PAPE).

## Appendix F. Additional details on the empirical application

Post-treatment estimation and standard errors. To prevent post-treatment information from entering the construction of the counterfactual outcomes, forecasts are generated recursively using only pre-treatment information. The August counterfactual is predicted using the two most recent pre-treatment outcomes. The September counterfactual is then obtained using the predicted August counterfactual, rather than the observed August outcome, together with the last available pre-treatment lag. Thus, realized post-treatment outcomes are never used as predictors of subsequent counterfactual outcomes. The bagging model is estimated using only pre-treatment unit-period observations for which the required outcome lags are available and employs 1,000 regression trees. Statistical uncertainty is quantified through a nonparametric bootstrap that repeats the entire estimation procedure, including model training, recursive forecasting, and aggregation of the unit-level effects. Standard errors are computed as the standard deviation of the resulting bootstrap estimates across replications.

Standard errors are computed from the cross-sectional dispersion of the unit-level forecast-based effects. For each target group $g$, the estimated effect is the average of $\hat{\tau}_{it} = Y_{it} - \hat{Y}_{it}(0,0)$, and its standard error is calculated as the sample standard deviation of the unit-level effects divided by the square root of the number of blocks in the group.

Table F1 reports the MLCM placebo estimates for the last pre-treatment period. These estimates are generally small and statistically insignificant, providing little evidence of systematic pre-treatment effects and thereby corroborating the credibility of the estimates reported in the main analysis.

**Table F1. Placebo effects for the last pre-treatment period**

| **Estimand** | **July 1 – July 17** |
|---|---|
| ATOT: directly protected blocks | -0.010 (0.025) |
| SE(1 block): one block away from treated units | 0.051* (0.030) |
| SE(2 blocks): two blocks away from treated units | -0.012 (0.017) |
| SE(>2 blocks): more than two blocks away from treated units | 0.010 (0.012) |
| PAPE: population average policy effect | 0.011 (0.010) |

*Notes: Standard errors are reported in parentheses. The placebo month is the first part of July 1994, from July 1 to July 17, immediately before the terrorist attack. The outcome is the number of car thefts normalized to a 30-day-equivalent monthly count. *, **, and *** denote statistical significance at the 10%, 5%, and 1% levels, respectively.*